\documentclass[12pt]{article}

\usepackage{subfig}
\usepackage{bm}
\usepackage{mathrsfs}
\usepackage{amsfonts}
\usepackage{srcltx}
\usepackage{graphicx}
\usepackage{amsmath}
\usepackage{amsthm}
\usepackage{titlesec}
\usepackage[authoryear]{natbib}
\usepackage{microtype}
\usepackage{enumerate}
\usepackage{xcolor}
\usepackage{here}
\usepackage{changes}
\usepackage{bbm}
\usepackage{multirow}
\usepackage{makecell}
\usepackage{tikz}
\usepackage{ulem}
\usepackage{pifont}
\usetikzlibrary{shapes}
\usetikzlibrary{positioning}
\usepackage{longtable,booktabs}
\usepackage{ulem}
\usepackage[left=2.5cm, right=2.5cm, top=2.5cm, bottom=2.5cm]{geometry}

\makeatletter
\def\singlespace{\def\baselinestretch{1}\@normalsize}

\def\0{\mbox{\boldmath$0$}}

\usepackage{framed}
\usepackage{mdframed}
\usepackage{graphicx}
\usepackage{float}

\usepackage{epstopdf}
\usepackage{amsmath}

\renewcommand{\baselinestretch}{1.3}
\newcommand{\doubleline}
{\addtolength{\baselineskip}{.4\baselineskip}}

\newtheoremstyle{highthm}{0mm plus .3ex minus .1ex}{3mm plus .3ex minus .1ex}{\slshape}{\parindent}{\bfseries\scshape}{}{4mm}{}
\theoremstyle{highthm}

\newcommand{\vs}{\vspace{0.1in}}

\makeatletter

\title{\bf Distributed Selective Inference for Quantile Regression}

\vspace{2cm}

\author{
Xiaohui Yuan\\
School of Mathematics and Statistics, Changchun University of Technology,\\
Changchun 130012, China\\
yuanxh@ccut.edu.cn\\
\\
Jiahan Teng\\
School of Mathematics and Statistics, Changchun University of Technology,\\
Changchun 130012, China\\
tengjh0714@163.com\\
\\
Yan Zhou\\
School of Mathematical Sciences, Institute of Statistical Sciences,\\ Shenzhen Key Laboratory of Advanced Machine Learning and Applications, \\Shenzhen University,\\
Shenzhen 518000, China\\
zhouy1016@szu.edu.cn\\
}
\date{}

\begin{document}
\maketitle

\doubleline {

\thispagestyle{empty}
\newpage
\thispagestyle{empty}
\begin{center}
ABSTRACT
\end{center}

We propose a distributed selective inference framework tailored for high-dimensional quantile regression. To enable valid post-selection inference in this context, we address the computational challenge posed by the non-smooth quantile loss via a response-surrogation strategy. This strategy transforms the problem into a penalized least-squares formulation, thereby facilitating distributed selective inference. For valid post-selection inference, a randomized procedure is introduced, in which the Lasso selection event is characterized through the associated Karush-Kuhn-Tucker conditions and the conditional distribution of the aggregated estimator is derived given the selection event. The resulting algorithm requires only three rounds of communication between local machines and the central server. Under standard regularity conditions, we establish the asymptotic validity of the proposed procedure and develop a large-deviation approximation to the selective likelihood for computationally tractable implementation. Simulation studies and a real-data application demonstrate the satisfactory finite-sample performance of the proposed method.
\vs

\pagestyle{plain}
\newpage
\pagenumbering{arabic}

\section{Introduction}

Modern high-dimensional data analysis routinely combines model selection with statistical inference. Sparse regularization methods, such as Lasso [28] and SCAD [8], simultaneously perform variable selection and parameter estimation. Conventional inference procedures that treat the selected model as fixed and
ignore the preceding model selection step can lead to distorted confidence
intervals and hypothesis tests when the same data are used for both selection
and inference [13,19]. Selective inference addresses this issue by conditioning explicitly on the realized selection event. For Lasso-regularized linear models, the Karush-Kuhn-Tucker (KKT) conditions characterize selection events through polyhedral constraints, leading to exact conditional distributions and valid selective p-values [18,29]. To improve statistical efficiency, randomized selective inference introduces additional randomization into the optimization problem and constructs tractable approximations to the selective likelihood while avoiding the information loss associated with data splitting [22,23,27]. Despite these advances, existing selective inference methodologies have primarily focused on mean regression models.

In many practical applications, datasets exhibit heavy-tailed distributions, outliers, and heterogeneous structures, which can substantially affect the robustness and reliability of statistical modeling and inference. Under such settings, conventional mean regression models can suffer from unstable estimation and reduced inferential accuracy due to their sensitivity to extreme observations [9,15]. Therefore, robust regression methods provide effective approaches for handling deviations from standard distributional assumptions
[9]. Among various robust regression approaches, quantile regression has emerged as an important framework for modeling conditional response quantiles and provides a robust alternative to mean regression[16,17]. High-dimensional quantile regression further incorporates sparse regularization to simultaneously achieve variable selection and estimation [2,7,25]. As variable selection becomes increasingly integrated into quantile regression, valid inference after data-driven selection has attracted growing attention. Several recent studies have investigated inference procedures for penalized quantile regression. Belloni et al. [3] developed score-based inference procedures for high-dimensional quantile regression without conditioning on selection events, while Wang et al. [31] proposed a randomized selective inference framework specifically designed for quantile regression models. These developments extend selective inference methodology to quantile settings. However, existing approaches assume centralized access to the full dataset.

Meanwhile, the rapid growth of large-scale datasets has led to increasingly distributed data storage and computation across multiple computing nodes due to storage limitations, privacy requirements, and data ownership constraints.
This has motivated extensive research on distributed statistical inference,
including divide-and-conquer methods, communication-efficient distributed
inference, and federated learning [1,4,12,14,20,32]. While distributed estimation has been well studied, most existing selective inference procedures remain inherently centralized. Recently, Liu and Panigrahi [21] proposed a distributed selective inference framework for generalized linear models, demonstrating that valid post-selection inference can be conducted without collecting the full dataset at a central server. Nevertheless, their framework is designed for smooth likelihood-based models and cannot be directly applied to quantile regression. To the best of our knowledge, distributed selective inference for high-dimensional quantile regression has not yet been investigated.

The contributions of this paper are summarized as follows.
First, we develop a distributed selective inference framework for high-dimensional quantile regression, extending selective inference beyond conventional mean regression models to robust quantile regression settings under distributed data environments.
Second, we perform quantile regression distributively across different machines and use surrogate responses to avoid the non-smooth quantile loss, thereby reducing the computational burden of high-dimensional quantile regression. The proposed procedure requires only three rounds of communication with a communication cost of order $O(d^2)$, independent of the ambient dimension $p$.
Finally, we establish the asymptotic validity of the proposed selective confidence intervals and hypothesis tests, and develop a large-deviation approximation for the selective likelihood. Simulation studies and real-data analysis further demonstrate the reliable finite-sample performance of the proposed method across different quantile levels.

The rest of this paper is organized as follows.
In Section 2, we introduce the distributed quantile regression framework and describe the surrogate response construction and randomized selection mechanism. We further derive the asymptotic distribution of the aggregated estimator under the distributed setting.
In Section 3, we develop the selective inference procedure for distributed quantile regression, including the construction of the selective conditional distribution and its large-deviation approximation.
In Sections 4 and 5, we present simulation studies and real-data applications, respectively.
Section 6 concludes the paper. All technical proofs are provided in the Appendix.

\section{Distributed Quantile Regression and Selective Inference Setup}

Let a vector  
\(\boldsymbol{v}=(v_1,\ldots,v_p)^\top\in\mathbb{R}^p\), and let
$
\|\boldsymbol{v}\|_1=\sum_{i=1}^p|v_i|,
\|\boldsymbol{v}\|_2=
\left(\sum_{i=1}^p v_i^2\right)^{1/2}
$
denote its \(\ell_1\)- and \(\ell_2\)-norms, respectively.
For a matrix
\(\boldsymbol{A}=(a_{ij})\in\mathbb{R}^{p\times q}\), we define
$
\|\boldsymbol{A}\|_\infty
=
\max_{1\leq i\leq p}
\sum_{j=1}^q|a_{ij}|
$
for its infinity norm.
Given index sets
\(S\subseteq\{1,\ldots,p\}\) and
\(T\subseteq\{1,\ldots,q\}\),
\(\boldsymbol{A}_{S,T}\) denotes the submatrix of
\(\boldsymbol{A}\) with rows indexed by \(S\) and columns indexed by \(T\).
For a symmetric matrix \(\boldsymbol{A}\), its smallest and largest
eigenvalues are denoted by
\(\Lambda_{\min}(\boldsymbol{A})\) and
\(\Lambda_{\max}(\boldsymbol{A})\).
The notation \(\operatorname{diag}(\cdot)\) represents a diagonal matrix.
For a vector \(\boldsymbol{\beta}\),
$
\operatorname{supp}(\boldsymbol{\beta})
=
\{j:\beta_j\neq0\}
$
denotes its support.

We consider a distributed quantile regression framework with one central machine and $K$ local machines. Let $Y\in\mathbb{R}$ denote the response variable, and let $\boldsymbol{X}=(X_1,\ldots,X_p)^{\top}\in\mathbb{R}^{p}$ denote the corresponding covariate vector. 
We consider the linear quantile regression model
\[
Q_{\tau}(Y\mid\boldsymbol{X})
=
\boldsymbol{X}^{\top}\boldsymbol{\beta}^{*},
\qquad \tau\in(0,1),
\]
where
\(\boldsymbol{\beta}^{*}
=
(\beta_1^*,\ldots,\beta_p^*)^{\top}
\in\mathbb{R}^{p}\)
is the true coefficient vector at the quantile level \(\tau\).
We assume that
$
\mathbb{E}(\boldsymbol{X})=\mathbf{0},
\operatorname{Var}(\boldsymbol{X})
=
\mathbb{E}
\left(
\boldsymbol{X}\boldsymbol{X}^{\top}
\right)
=
\boldsymbol{\Sigma}.
$
Let
$
\varepsilon
=
Y-\boldsymbol{X}^{\top}\boldsymbol{\beta}^{*}
$
denote the regression error. We assume that \(\varepsilon\) is independent of
\(\boldsymbol{X}\) and follows an unknown distribution with density function
\(f(\cdot)\). The error distribution is allowed to be heavy-tailed, and its variance may even be infinite. 
Under these assumptions, the true regression coefficient is characterized as the minimizer of the population quantile loss
\[
\boldsymbol{\beta}^{*}
=
\arg\min_{\boldsymbol{\beta}\in\mathbb{R}^{p}}
\mathbb{E}
\left[
\rho_{\tau}
\left(
Y-\boldsymbol{X}^{\top}\boldsymbol{\beta}
\right)
\right],
\]
where
$\rho_{\tau}(u) = u\left(\tau - \mathbb{I}\{u \le 0\}\right)$
is the quantile check loss function, \(\mathbb{I}(\cdot)\) is the indicator function.

Suppose that $n$ independent and identically distributed observations
$\{(Y_i,\boldsymbol{X}_i), i\in[n]\}$
are distributed across one central machine and $K$ local machines, where $[n]=\{1,\ldots,n\}$. Let $\mathcal{C}^{(k)}\subset[n]$ denote the index set of observations stored on machine $k$, where $k\in\{0\}\cup[K]$. The index sets are mutually disjoint and satisfy
$
\bigcup_{k=0}^{K}\mathcal{C}^{(k)}=[n].
$

The dataset stored on machine $k$ is denoted by$
D^{(k)}=\{(Y_i,\boldsymbol{X}_i):i\in\mathcal{C}^{(k)}\}$.
Let $n_k=|\mathcal{C}^{(k)}|$ and $\alpha_k=n_k/n$. Then
$
n=\sum_{k=0}^{K}n_k,$
$
\sum_{k=0}^{K}\alpha_k=1 .
$
For observations stored on machine $k$, the model can be written as
\[
Q_{\tau}(Y_i\mid\boldsymbol{X_i})
=
\boldsymbol{X_i}^{\top}\boldsymbol{\beta}^{*},
\qquad
i\in\mathcal{C}^{(k)}.
\]

Since the quantile loss is non-smooth, directly developing a distributed selective inference procedure is computationally prohibitive. To facilitate subsequent inference, we adopt the response surrogation technique proposed by Chen et al. (2020) which reformulates the quantile regression problem as a penalized least-squares problem.
Specifically, we first compute the initial penalized quantile regression estimator
\[
\tilde{\boldsymbol{\beta}}^{(k)}
=
\arg\min_{\boldsymbol{\beta}\in\mathbb{R}^{p}}
\frac1{n_k}
\sum_{i\in\mathcal{C}^{(k)}}
\rho_{\tau}
\left(
Y_i-\boldsymbol{X}_i^{\top}\boldsymbol{\beta}
\right)
+\lambda_0\|\boldsymbol{\beta}\|_1,
\]
where \(\lambda_0\) is the initial quantile Lasso regularization parameter.
The surrogate response is constructed as
\[
\tilde{Y}_i
=
\boldsymbol{X}_i^{\top}\tilde{\boldsymbol{\beta}}^{(k)}
-
\frac1{\widehat{f}^{(k)}(0)}
\left\{
\mathbb{I}
\left(
Y_i-\boldsymbol{X}_i^{\top}\tilde{\boldsymbol{\beta}}^{(k)}
\le0
\right)
-\tau
\right\},
\qquad
i\in\mathcal{C}^{(k)},
\]
with the density of the random error at zero is estimated by
\[
\widehat{f}^{(k)}(0)
=
\frac1{n_kh}
\sum_{i\in\mathcal{C}^{(k)}}
G
\left(
\frac{
Y_i-\boldsymbol{X}_i^{\top}\tilde{\boldsymbol{\beta}}^{(k)}
}{h}
\right).
\]
Here, \(h>0\) is the bandwidth parameter, \(G(\cdot)\) is a kernel function.
Based on \(\tilde{\boldsymbol{\beta}}^{(k)}\) and \(\widehat{f}^{(k)}(0)\),
the corresponding surrogate estimator is obtained by solving the penalized least-squares problem
\[
\widehat{\boldsymbol{\beta}}^{\lambda,(k)}
=
\arg\min_{\boldsymbol{\beta}\in\mathbb{R}^{p}}
\frac1{n_k}
\sum_{i\in\mathcal{C}^{(k)}}
\left(
\tilde{Y}_i
-
\boldsymbol{X}_i^{\top}\boldsymbol{\beta}
\right)^2
+
\lambda\|\boldsymbol{\beta}\|_1,
\tag{1}
\]
where \(\lambda\) is the surrogate Lasso penalty parameter.

The response surrogation approach has subsequently been adopted in a variety of quantile regression settings, including distributed estimation, varying coefficient models, and functional quantile regression \cite{Chen2026,Jiang2022,Jin2024}.
Based on the surrogate Lasso estimator in (1), each local machine performs variable selection independently. Let
\[
\widehat{E}^{(k)}
=
\left\{
j\in[p]:
\widehat{\boldsymbol{\beta}}_j^{\lambda,(k)}\neq0
\right\},
\]
represent the set of predictors with non-zero lasso coefficients selected at machine \(k\) for \(1\le k\le K\). We denote the realized value of \(\widehat{E}^{(k)}\) by \(E^{(k)}\), and let \(d^{(k)}=|E^{(k)}|\) denote the number of selected variables on machine \(k\).
The selected variable sets are transmitted to the central machine and aggregated by taking their union,
\[
\widehat{E}
=
\bigcup_{k\in[K]}
\widehat{E}^{(k)}.
\]
We let \( E \) be the observed value of \( \widehat{E} \) and let \( d = |E| \).

To characterize the resulting selection event, we adopt the randomized Lasso framework of Tian et al. (2016). The surrogate Lasso problem on machine \(k\) can be equivalently written as
\[
\widehat{\boldsymbol{\beta}}^{\lambda,(k)}
=
\arg\min_{\boldsymbol{\beta}\in\mathbb{R}^p}
\frac1n
\sum_{k=0}^{K}
\sum_{i\in\mathcal{C}^{(k)}}
\left(
\widetilde{Y}_i-
\boldsymbol{X}_i^\top\boldsymbol{\beta}
\right)^2
+
\lambda\|
\boldsymbol{\beta}
\|_1
-
\boldsymbol{\omega}^{(k)\top}\boldsymbol{\beta},
\tag{2}
\]
where the randomization variable is defined as
\[
\boldsymbol{\omega}^{(k)}
=
\frac1n
\sum_{k=0}^{K}
\sum_{i\in\mathcal{C}^{(k)}}
\boldsymbol{X}_i
\left(
\boldsymbol{X}_i^\top
\widehat{\boldsymbol{\beta}}^{\lambda,(k)}
-
\widetilde{Y}_i
\right)
-
\frac1{n_k}
\sum_{i\in\mathcal{C}^{(k)}}
\boldsymbol{X}_i
\left(
\boldsymbol{X}_i^\top
\widehat{\boldsymbol{\beta}}^{\lambda,(k)}
-
\widetilde{Y}_i
\right).
\]

The KKT conditions of problem (1) and  (2)
are the same. Note that the KKT conditions are given as follows:
\begin{gather*}
\frac1n
\sum_{k=0}^{K}
\sum_{i\in\mathcal{C}^{(k)}}
\boldsymbol{X}_i
\left(
\boldsymbol{X}_i^\top
\widehat{\boldsymbol{\beta}}^{\lambda,(k)}
-
\widetilde{Y}_i
\right)
+
\boldsymbol{\gamma}^{(k)}
=
\boldsymbol{\omega}^{(k)},\\
\boldsymbol{\gamma}_{E^{(k)}}^{(k)}
=
\boldsymbol{\lambda}\boldsymbol{s}^{(k)},
\qquad
\boldsymbol{s}^{(k)}
=
\operatorname{sign}
\left(
\widehat{\boldsymbol{\beta}}_{E^{(k)}}^{\lambda,(k)}
\right),
\\
\boldsymbol{\gamma}_{-E^{(k)}}^{(k)}
=
\boldsymbol{\lambda}\boldsymbol{z}^{(k)},
\qquad
\|
\boldsymbol{z}^{(k)}
\|_\infty
\le1,
\end{gather*}
where  $\boldsymbol{\gamma}^{(k)}$ denotes the subgradient of the
$\ell_1$-penalty evaluated at the Lasso solution. The operator
$\operatorname{sign}(\cdot)$ denotes the componentwise sign function.
Accordingly, $\boldsymbol{s}^{(k)}$
records the signs of the active coefficients, while
$\boldsymbol{z}^{(k)}$ denotes the subgradient corresponding to the
inactive coordinates.

Let
$
\bar d=\sum_{k\in[K]}d^{(k)}.
$
For each machine \(k\), let
$
\boldsymbol{b}^{(k)}
=
\widehat{\boldsymbol{\beta}}_{E^{(k)}}^{\lambda,(k)}
\in\mathbb{R}^{d^{(k)}}
$
denote the active-set Lasso estimator, and let
\(\widehat{\boldsymbol{B}}^{(k)}\)
be the corresponding random vector. Stacking
\(\widehat{\boldsymbol{B}}^{(k)}\)
across all local machines yields
$
\widehat{\boldsymbol{B}}
\in
\mathbb{R}^{\bar d},
$
whose realized value is denoted by
\(\boldsymbol{b}\). 

Our procedure conditions on the event
\[
\left\{
\widehat{\boldsymbol{\Gamma}}^{(k)}
=
\boldsymbol{\gamma}^{(k)},
\quad
\forall k\in[K]
\right\},
\tag{3}
\]
where
\(\widehat{\boldsymbol{\Gamma}}^{(k)}\)
denotes the subgradient of the Lasso penalty on machine \(k\), with realized
value \(\boldsymbol{\gamma}^{(k)}\).
The event in (3) is contained in
\[
\left\{
\widehat{E}^{(k)}=E^{(k)},
\quad
\forall k\in[K]
\right\},
\]
and therefore is also contained in the global selection event
\(\{\widehat{E}=E\}\).

Let
\(\widehat{\boldsymbol{S}}\in\mathbb{R}^{\bar d}\)
and
\(\widehat{\boldsymbol{Z}}\in\mathbb{R}^{pK-\bar d}\)
denote the stacked signs of the active Lasso coefficients and the stacked
subgradients corresponding to the inactive variables, respectively, with realized values
\(\boldsymbol{s}\) and \(\boldsymbol{z}\).
By the KKT conditions, the conditioning event in (3) can be equivalently
expressed as
\[
\left\{
\operatorname{sign}(\widehat{\boldsymbol{B}})
=
\boldsymbol{s},
\quad
\widehat{\boldsymbol{Z}}
=
\boldsymbol{z}
\right\}.
\]
Hereafter, we refer to the conditioning event in (3) as the selection event.
For the selected model $E$, let
$\boldsymbol{\beta}_E^*\in\mathbb R^d$
denote the subvector of the true coefficient vector
$\boldsymbol{\beta}^*$ indexed by $E$.
Let
$
\mathcal A^*
=
\left\{
j\in[p]:\beta_j^*\neq0
\right\}
$
denote the true support of $\boldsymbol{\beta}^*$.
Throughout the asymptotic analysis, we assume that
$
\mathcal A^*\subseteq E.
$
Consequently,
$\boldsymbol{\beta}_{-E}^*=\boldsymbol 0$, and hence
$
\boldsymbol X_i^\top\boldsymbol{\beta}^*
=
\boldsymbol X_{i,E}^\top\boldsymbol{\beta}_E^*.
$

To establish the asymptotic properties of the local and aggregated
estimators, we impose the following regularity conditions on the error
distribution, covariates, kernel function, and selected models.

\textbf{Assumption 1}.
The density function of the noise \(f(\cdot)\) is bounded and Lipschitz continuous. Moreover, assume that \(f(0)>c\) for some constant \(c>0\).

\textbf{Assumption 2}.
Assume that the covariate vector \(\boldsymbol X\) satisfies the sub-Gaussian condition
\[
\sup_{\|\boldsymbol\theta\|_2=1}
\mathbb E
\exp
\left(
m(\boldsymbol\theta^\top\boldsymbol X)^2
\right)
\le C,
\]
for some constants \(m>0\) and \(C>0\).

\textbf{Assumption 3}.
Suppose that \(\boldsymbol{\Sigma}\) satisfies
\[
\left\|
\boldsymbol{\Sigma}_{-E^{(k)}, E^{(k)}}
\boldsymbol{\Sigma}_{E^{(k)}, E^{(k)}}^{-1}
\right\|_\infty
\le
1-\eta_k,
\]
for some \(0<\eta_k<1\) and all $k$. In addition, assume
$
c_0^{-1}
\le
\Lambda_{\min}(\boldsymbol{\Sigma})
\le
\Lambda_{\max}(\boldsymbol{\Sigma})
\le
c_0,
$
for some constant \(c_0>0\).

\textbf{Assumption 4}.
Assume that the kernel function \(G(\cdot)\) is integrable with
$
\int_{-\infty}^{\infty}G(u)\,du=1,
$
and satisfies \(G(u)=0\) whenever \(|u|\ge1\). Furthermore, \(G(\cdot)\) is differentiable with a bounded derivative.

\textbf{Assumption 5}.
For each \(k\in[K]\), let
\(
\widetilde E^{(k)}
=
E\setminus E^{(k)}.
\)
For every
\(j\in\widetilde E^{(k)}\),
either
\(
\beta_{E,j}^*
=
O(n^{-1/2})
\)
or
\[
\boldsymbol X_{E^{(k)}}^\top
\boldsymbol\Sigma_{E^{(k)},E^{(k)}}^{-1}
\boldsymbol\Sigma_{E^{(k)},j}
=
\boldsymbol X_j.
\]

Assumptions 1-4 follow the standard regularity conditions
in Chen et al. (2020). Specifically, Assumption 1 imposes smoothness
and positivity conditions on the error density, Assumption 2 controls the tail behavior of the covariates through a sub-Gaussian condition, Assumption 3 corresponds to the standard
irrepresentable condition for support recovery, Assumption 4
specifies regularity of the kernel function.
Assumption 5, following Liu and Panigrahi (2025), further characterizes
predictors that belong to the aggregated model but are not selected on a
particular local machine.

Under the above assumptions, the first result establishes an asymptotically linear representation of the local estimator. Conditional on the selection event, we construct the estimator for distributed selective inference. The local estimator on machine \(k\) is defined as
\[
\widehat{\boldsymbol{\beta}}_E^{(k)}
=
\arg\min_{\boldsymbol{\beta}\in\mathbb{R}^{d}}
\frac1{n_k}
\sum_{i\in\mathcal{C}^{(k)}}
\left(
\widetilde{Y}_i-
\boldsymbol{X}_{i,E}^{\top}\boldsymbol{\beta}
\right)^2.
\]

\textbf{Proposition 1}.
Under assumptions 1-4, for each $k\in\{0\}\cup[K]$,
the local estimator \(\widehat{\boldsymbol{\beta}}_E^{(k)}\) satisfies
\[
\sqrt{n_k}
\left(
\widehat{\boldsymbol{\beta}}_E^{(k)}
-
\boldsymbol{\beta}_E^*
\right)
=
-\frac{1}{f(0)}
\boldsymbol{\Sigma}_{E,E}^{-1}
\frac{1}{\sqrt{n_k}}
\sum_{i\in\mathcal{C}^{(k)}}
\boldsymbol{X}_{i,E}
\left\{
\mathbb{I}(\varepsilon_i\leq0)-\tau
\right\}
+
o_p(1).
\]

The proof of Proposition 1 is provided in Appendix A.1. 

The local estimators are finally combined through the weighted aggregation rule
\[
\widehat{\boldsymbol{\beta}}_E
=
\left(
\sum_{k=0}^{K}
\alpha_k
\widehat{\boldsymbol{\Sigma}}_{E,E}^{(k)}
\right)^{-1}
\left(
\sum_{k=0}^{K}
\alpha_k
\widehat{\boldsymbol{\Sigma}}_{E,E}^{(k)}
\widehat{\boldsymbol{\beta}}_E^{(k)}
\right),
\tag{4}
\]
where
$
\widehat{\boldsymbol{\Sigma}}_{E,E}^{(k)}
=
\frac{1}{n_k}
\sum_{i\in\mathcal{C}^{(k)}}
\boldsymbol{X}_{i,E}
\boldsymbol{X}_{i,E}^{\top}.
$

\textbf{Proposition 2}.
Under assumptions 1-4, the aggregated estimator $\widehat{\boldsymbol{\beta}}_E$ satisfies
\[
\sqrt n
\left(
\widehat{\boldsymbol\beta}_E
-
\boldsymbol\beta_E^*
\right)
=
-\frac1{f(0)}
\boldsymbol\Sigma_{E,E}^{-1}
\frac1{\sqrt n}
\sum_{i\in[n]}
\boldsymbol X_{i,E}
\left\{
\mathbb I(\varepsilon_i\le0)-\tau
\right\}
+
o_p(1).
\]

The proof of Proposition 2 is deferred to Appendix A.2. 

The local and aggregated estimators admit an asymptotically linear representation, which serves as the basis for establishing its limiting distribution. We next derive the asymptotic distribution of the randomization variables.

\textbf{Proposition 3}.
Suppose that
\(n_k/n\rightarrow\alpha_k\)
for
\(k\in[K]\)
and
\(
\sum_{k=1}^K\alpha_k<1.
\)
Let \( \boldsymbol\Omega = (\boldsymbol\omega^{(1),\top}, \ldots, \boldsymbol\omega^{(K),\top})^{\top} \in \mathbb{R}^{Kp} \) be the stack of the \( K \) randomization variables.
Let
$
\boldsymbol U
=
\operatorname{diag}
(\alpha_1^{-1},\ldots,\alpha_K^{-1})
-
\mathbf1_{K\times K},
$
where $\mathbf1_{K\times K}$ denotes the $K\times K$ matrix with all entries equal
to $\mathbf1$.
Define
\[
\boldsymbol V_\Omega
=
\frac{\tau(1-\tau)}{f^2(0)}
\boldsymbol U\otimes\boldsymbol\Sigma.
\]
Under Assumptions 1-5,
\[
\sqrt n\,\boldsymbol\Omega
\stackrel d\Longrightarrow
\mathcal{N}_{pK}
(
\mathbf0,
\boldsymbol V_\Omega
),
\qquad
n\rightarrow\infty.
\]

The proof of Proposition 3 is provided in Appendix A.3. 

To derive the joint limiting distribution required for selective inference, it is necessary to account for the information contained in variables outside the selected model. For this purpose, we introduce the auxiliary statistic
\[
\widehat{\boldsymbol\beta}_{-E}^{\perp}
=
\frac1n
\sum_{k=0}^{K}
\sum_{i\in\mathcal C^{(k)}}
\boldsymbol X_{i,-E}^{\top}
\left(
\boldsymbol X_{i,E}^{\top}
\widehat{\boldsymbol\beta}_E
-
\widetilde Y_i
\right).
\]

Combining the above results, we obtain the joint asymptotic distribution of
\(
\widehat{\boldsymbol\beta}_E,
\widehat{\boldsymbol\beta}_{-E}^{\perp}
\)
and
\(
\boldsymbol\Omega.
\)

\textbf{Theorem 1}. Under Assumptions 1-5, it holds that
\[
\sqrt{n}\begin{pmatrix}
\widehat{\boldsymbol{\beta}}_E - \boldsymbol{\beta}_E^* \\
\widehat{\boldsymbol{\beta}}_{-E}^\perp \\
\boldsymbol{\Omega}
\end{pmatrix}
\stackrel{d}{\Rightarrow}
\mathcal{N}\left(
\begin{pmatrix}
\mathbf{0} \\ \mathbf{0} \\ \mathbf{0}
\end{pmatrix},\
\begin{pmatrix}
\boldsymbol{D}_{E, E} & \mathbf{0} & \mathbf{0} \\
\mathbf{0} & \boldsymbol{D}/\boldsymbol{D}_{E, E} & \mathbf{0} \\
\mathbf{0} & \mathbf{0} & \boldsymbol V_\Omega
\end{pmatrix}
\right),
\]
here, $
\boldsymbol{D}
=
\frac{\tau(1-\tau)}{f^2(0)}
\boldsymbol{\Sigma}^{-1},
$ $\boldsymbol{\Sigma}=\begin{pmatrix}\boldsymbol{\Sigma}_{E,E}&\boldsymbol{\Sigma}_{E,-E}\\\boldsymbol{\Sigma}_{-E,E}&\boldsymbol{\Sigma}_{-E,-E}\end{pmatrix}$,
\(\boldsymbol{D}_{E,E}\) denotes the submatrix of
\(\boldsymbol{D}\) corresponding to the selected variables \(E\), $\boldsymbol{D}/\boldsymbol{D}_{E,E}= \frac{\tau(1-\tau)}{f^2(0)}
    \Big(\boldsymbol{\Sigma}_{-E,-E}-\boldsymbol{\Sigma}_{-E,E}\boldsymbol{\Sigma}_{E,E}^{-1}\boldsymbol{\Sigma}_{E,-E}\Big).
$

The proof of Theorem 1 is deferred to Appendix A.4. Theorem 1 shows that
\(\widehat{\boldsymbol\beta}_{-E}^{\perp}\)
is an ancillary statistic for inference on the selected model. Moreover, the randomization variables
\(\boldsymbol\Omega\)
are asymptotically independent of
\(
\widehat{\boldsymbol\beta}_E
\)
and
\(
\widehat{\boldsymbol\beta}_{-E}^{\perp},
\)
as indicated by the block-diagonal structure of the limiting covariance matrix.

\section{Selective Inference for Distributed Quantile Regression}

In this section, we derive the selective distribution of the aggregated estimator, which forms the basis for valid post-selection inference. Specifically, the selective distribution is obtained by conditioning the asymptotic distribution established in Theorem 1 on the selection event defined above.

To simplify the presentation of the selective likelihood, we first introduce several quantities and matrices that arise in the limiting distribution. By Theorem 1, the joint distribution of
\((\widehat{\boldsymbol{\beta}}_E,\widehat{\boldsymbol{\beta}}_{-E}^{\perp},\boldsymbol{\Omega})\)
converges weakly to the corresponding Gaussian distribution.

For \(j,k\in[K]\), define
$
\boldsymbol{g}_k^{(j)}
=
\boldsymbol{\gamma}_{E^{(k)}}^{(j)}
\in
\mathbb{R}^{d_k},
$
where
\(\boldsymbol{g}_k^{(j)}\)
collects the coordinates of
\(\boldsymbol{\gamma}^{(j)}\)
corresponding to the variables in
\(E^{(k)}\).
Similarly, define
$
\boldsymbol{g}^{(j)}
=
\boldsymbol{\gamma}_{E}^{(j)}
\in
\mathbb{R}^{d},
$
which \(\boldsymbol{g}^{(j)}\) consists of the coordinates of
\(\boldsymbol{\gamma}^{(j)}\)
corresponding to the aggregated model \(E\).
Let
$
\alpha_0
=
1-\sum_{k\in[K]}\alpha_k.
$
Based on these quantities, define the matrices
\[
\boldsymbol{\Xi}\in\mathbb{R}^{\bar d\times\bar d},
\quad
\boldsymbol{\Psi}\in\mathbb{R}^{\bar d\times d},
\quad
\boldsymbol t\in\mathbb{R}^{\bar d},
\quad
\boldsymbol{\Theta}\in\mathbb{R}^{d\times d},
\quad
\boldsymbol{\Pi}\in\mathbb{R}^{d\times d},
\quad
\boldsymbol{\kappa}\in\mathbb{R}^{d},
\]
where the \((j,k)\)-th block of
\(\boldsymbol{\Xi}^{-1}\)
is
\[
\left\{
\boldsymbol{\Xi}^{-1}
\right\}_{j,k}
=
\begin{cases}
\left(
\alpha_k+\dfrac{\alpha_k^2}{\alpha_0}
\right)
\boldsymbol{\Sigma}_{E^{(k)},E^{(k)}},
&
j=k,
\\[0.3cm]
\dfrac{\alpha_j\alpha_k}{\alpha_0}
\boldsymbol{\Sigma}_{E^{(j)},E^{(k)}},
&
j\neq k.
\end{cases}
\]
The \((k,1)\)-th blocks of
\(\boldsymbol{\Xi}^{-1}\boldsymbol{\Psi}\)
and
\(\boldsymbol{\Xi}^{-1}\boldsymbol t\)
are given by
\[
\left\{
\boldsymbol{\Xi}^{-1}\boldsymbol{\Psi}
\right\}_k
=
\frac{\alpha_k}{\alpha_0}
\boldsymbol{\Sigma}_{E^{(k)},E},
\qquad
\left\{
\boldsymbol{\Xi}^{-1}\boldsymbol t
\right\}_k
=
-\alpha_k\boldsymbol g_k^{(k)}
-
\frac{\alpha_k}{\alpha_0}
\sum_{j=1}^{K}
\alpha_j
\boldsymbol g_k^{(j)}.
\]
Let
\[
\boldsymbol{\Theta}^{-1}
=
\frac1{\alpha_0}
\boldsymbol D_{E,E}^{-1}
-
\boldsymbol{\Psi}^{\top}
\boldsymbol{\Xi}^{-1}
\boldsymbol{\Psi},
\]
\[
\boldsymbol{\Theta}^{-1}\boldsymbol{\Pi}
=
\boldsymbol D_{E,E}^{-1},
\]
\[
\boldsymbol{\Theta}^{-1}\boldsymbol{\kappa}
=
\boldsymbol{\Psi}^{\top}
\boldsymbol{\Xi}^{-1}
\boldsymbol t
+
\sum_{j=1}^{K}
\frac{\alpha_j}{\alpha_0}
\boldsymbol g^{(j)}.
\]

With these definitions, we are now ready to derive the selective likelihood of the aggregated estimator.

\textbf{Theorem 2}. Suppose that the conditions stated in Theorem 1 hold. The selective likelihood function, based on the asymptotic distribution of \(\sqrt{n}\widehat{\boldsymbol{\beta}}_E \,\big|\, \big\{\widehat{\boldsymbol{\Gamma}}^{(k)} = \boldsymbol{\gamma}^{(k)},\ \forall k \in [K]\big\}\), is equal to \[ f \big(\boldsymbol{\beta}_E; \widehat{\boldsymbol{\beta}}_E, \boldsymbol{s}, \boldsymbol{z}\big) = \frac{\varphi\big(\sqrt{n}\widehat{\boldsymbol{\beta}}_E;\, \boldsymbol{\Pi}\sqrt{n}\boldsymbol{\beta}_E + \boldsymbol{\kappa},\, \boldsymbol{\Theta}\big)} {\mathbb{P}\Big[\sqrt{n}\widehat{\boldsymbol{B}} \in \mathcal{O} \,\big|\, \widehat{\boldsymbol{Z}} = \boldsymbol{z}\Big]}, \] where \[ \begin{aligned} \mathbb{P}\Big[\sqrt{n}\widehat{\boldsymbol{B}} \in \mathcal{O} \,\big|\, \widehat{\boldsymbol{Z}} = \boldsymbol{z}\Big] &= \int \varphi\big(\sqrt{n}\widehat{\boldsymbol{\beta}}_E;\, \boldsymbol{\Pi}\sqrt{n}\boldsymbol{\beta}_E + \boldsymbol{\kappa},\, \boldsymbol{\Theta}\big) \\ &\quad \cdot \varphi\big(\sqrt{n}\widehat{\boldsymbol{B}};\, \boldsymbol{\Psi}\sqrt{n}\widehat{\boldsymbol{\beta}}_E + \boldsymbol{t},\, \boldsymbol{\Xi}\big) \cdot \mathbf{1}\big\{\widehat{\boldsymbol{B}} \in \mathcal{O}\big\} \,\mathrm{d}\widehat{\boldsymbol{\beta}}_E \mathrm{d}\widehat{\boldsymbol{B}},
\end{aligned} \] 
with \(\varphi(\cdot;\boldsymbol\mu,\boldsymbol\Sigma)\)
denote the density of the multivariate normal distribution with mean
\(\boldsymbol\mu\)
and covariance matrix
\(\boldsymbol\Sigma\).
Here,
$
\mathcal O
=
\left\{
\boldsymbol v\in\mathbb R^{\bar d}:
\operatorname{sign}(\boldsymbol v)
=
\operatorname{sign}(\boldsymbol s)
\right\},
$
which denotes the orthant determined by the observed sign vector
\(\boldsymbol s\).

The proof is deferred to the Appendix A.5. 

The selective likelihood derived above provides the foundation for constructing selective confidence intervals and hypothesis tests for the selected variables.

The proposed inference procedure can be implemented in a communication-efficient manner.
The selective inference framework only requires the exchange of low-dimensional summary statistics between the local machines and the central machine.
Specifically, the procedure involves three communication rounds: local machines first transmit their selected models, the central machine aggregates these models and returns the selected set, and local machines subsequently send the required summary statistics for inference.
The complete communication procedure is summarized in Algorithm 1.

\begin{center}
\begin{tabular}{p{14cm}}
\toprule
\textbf{Algorithm 1: Communication procedure}\\
\midrule

\textbf{Step 1. Variable selection at local machines}\\[3pt]
Machine \(k\) solves Problem (1) and transmits
\(E^{(k)}=\operatorname{Supp}(\widehat{\boldsymbol{\beta}}^{\lambda,(k)})\)
to the central machine.\\[6pt]

\textbf{Step 2. Model aggregation and broadcast}\\[3pt]
The central machine constructs
\(
E=\bigcup_{k\in[K]}E^{(k)}
\)
and broadcasts the aggregated model to all local machines.\\[6pt]

\textbf{Step 3. Transmission of summary statistics}\\[3pt]
Each local machine sends
$
\widehat{\boldsymbol{\beta}}_E^{(k)},
\widehat{\boldsymbol{\Sigma}}_{E,E}^{(k)},
\boldsymbol{\gamma}_E^{(k)}
$
to the central machine.
\\
\bottomrule
\end{tabular}
\end{center}

The communication procedure described in Algorithm 1 requires only three rounds of communication. The first two rounds involve only the indices of the selected variables, which are used to construct the aggregated model \(E\).

In the third communication round, each local machine transmits the local estimator \(\widehat{\boldsymbol{\beta}}_E^{(k)}\), the local observed information matrix \(\widehat{\boldsymbol{\Sigma}}_{E,E}^{(k)}\), and the corresponding subgradient \(\boldsymbol{\gamma}_E^{(k)}\) to the central machine. The local estimator and the observed information matrix are required to construct the aggregated estimator and the aggregated information matrix in (4). In addition, the subgradient variables appear explicitly in the selective likelihood derived in Theorem 2 and are therefore required to account for the model selection event. Hence, the central machine needs the subgradient vectors \(\boldsymbol{\gamma}^{(j)}\) restricted to the indices in \(E\cup\bigcup_{k\in[K]}E^{(k)}\). Since we assume
$
E=\bigcup_{k\in[K]}E^{(k)},
$
machine \(j\) only needs to transmit \(\boldsymbol{\gamma}_E^{(j)}\).

Since \(\widehat{\boldsymbol{\Sigma}}_{E,E}^{(k)}\) is a \(d\times d\) matrix, the communication cost for each local machine is of order \(O(d^2)\), which is independent of the ambient dimension \(p\). Consequently, as long as the selected model remains sparse, the proposed framework achieves communication-efficient selective inference. Extensions to more general aggregation rules are provided in the Appendix.

To perform selective inference, it remains to evaluate the selective likelihood established in Theorem 2. Taking the logarithm of the selective likelihood yields
\[
\log \varphi \left( \sqrt{n} \widehat{\boldsymbol{\beta}}_E;
\boldsymbol{\Pi}\sqrt{n}\boldsymbol{\beta}_E+\boldsymbol{\kappa},
\boldsymbol{\Theta}
\right)
-
\log
\mathbb{P}
\left[
\sqrt{n}\widehat{\boldsymbol{B}}\in\mathcal{O}
\,\big|\,
\widehat{\boldsymbol{Z}}=\boldsymbol{z}
\right].
\tag{5}
\]

The first term is the log-density of a multivariate normal distribution and can be evaluated directly. The computational difficulty lies in the second term, which involves the probability of the selection event. Following the large-deviation approach of Panigrahi and Taylor (2023), we approximate this probability by its corresponding large-deviation limit. The approximation is developed under the following asymptotic regime and regularity conditions.

Let \(\{a_n\}\) be a positive sequence satisfying
$
a_n\rightarrow\infty,
a_n=o(\sqrt n).
$
Assume that
\(\boldsymbol{\beta}_E=\boldsymbol{\beta}_{E,n}\)
satisfies
$
\sqrt n\,\boldsymbol{\beta}_{E,n}
=
a_n\boldsymbol{\beta}_E^*
\in\mathbb{R}^{|E|},
$
where
\(\boldsymbol{\beta}_E^*\)
does not depend on \(n\).
Under this asymptotic regime, the proof of Theorem 1 implies the following asymptotic linear representation:
\[
\sqrt{n}
\begin{pmatrix}
\widehat{\boldsymbol{\beta}}_E-\boldsymbol{\beta}_{E,n}\\
\widehat{\boldsymbol{\beta}}_{-E}^{\perp}\\
\boldsymbol{\Omega}
\end{pmatrix}
=
\sqrt n\,\bar{\boldsymbol E}_n
+
\boldsymbol R_n,
\tag{6}
\]
where
$
\bar{\boldsymbol E}_n
=
\frac1n
\sum_{i=1}^n
\boldsymbol e_{i,n}
$
is the average of \(n\) independent and identically distributed random vectors, and
\(\boldsymbol R_n=o_p(1)\).

\textbf{Assumption 6}.
Assume that
\[
\mathbb E
\left[
\exp\!\left(
\lambda\|\boldsymbol e_{1,n}\|_2
\right)
\right]
<
\infty
\]
for some \(\lambda \in \mathbb{R}^+\),
and
$
\lim_{n\rightarrow\infty}
\frac1{a_n^2}
\log
\mathbb P
\left[
\frac1{a_n}
\|\boldsymbol R_n\|_2
>
\epsilon
\right]
=
-\infty
$
for every
\(\epsilon>0\),
where
\(\boldsymbol e_{i,n}\)
and
\(\boldsymbol R_n\)
are defined in (6).

Assumption 6 guarantees that the sample average \(\bar{\boldsymbol E}_n\) satisfies the regularity conditions required for the large-deviation approximation. The second condition ensures that the remainder term \(\boldsymbol R_n\) is asymptotically negligible on the large-deviation scale.

The next assumption guarantees that small perturbations do not affect the limiting large-deviation probability.

\textbf{Assumption 7}.
For a fixed convex set
\(\mathcal R_0\subseteq\mathbb R^{p(K+1)}\),
and any
\(\boldsymbol O=O_p(1)\),
assume that
\[
\lim_{n\to\infty}
\frac1{a_n^2}
\left\{
\log
\mathbb P
\left[
\frac1{a_n}
\begin{pmatrix}
\sqrt n\,\widehat{\boldsymbol\beta}_E\\
\sqrt n\,\widehat{\boldsymbol\beta}_{-E}^{\perp}\\
\sqrt n\,\boldsymbol\Omega
\end{pmatrix}
\in
\mathcal R_0
\right]
-
\log
\mathbb P
\left[
\frac1{a_n}
\begin{pmatrix}
\sqrt n\,\widehat{\boldsymbol\beta}_E\\
\sqrt n\,\widehat{\boldsymbol\beta}_{-E}^{\perp}\\
\sqrt n\,\boldsymbol\Omega
\end{pmatrix}
+
\frac1{a_n}\boldsymbol O
\in
\mathcal R_0
\right]
\right\}
=
0.
\]

The probability of the selection event can be expressed in terms of
\(\widehat{\boldsymbol{\beta}}_E\),
\(\widehat{\boldsymbol{\beta}}_{-E}^{\perp}\),
and
\(\boldsymbol{\Omega}\)
as
\[
\mathbb P
\left[
\frac1{a_n}
\begin{pmatrix}
\sqrt n\,\widehat{\boldsymbol\beta}_E\\
\sqrt n\,\widehat{\boldsymbol\beta}_{-E}^{\perp}\\
\sqrt n\,\boldsymbol\Omega
\end{pmatrix}
+
\frac1{a_n}\boldsymbol O
\in
\mathcal R_0
\right].
\]

Under Assumptions 6 and 7, the logarithm of the normalizing constant admits the following large-deviation approximation.

\textbf{Theorem 3}.
Under Assumptions 6 and 7, define
\[
\begin{aligned}
L_n=\inf_{\boldsymbol b,\boldsymbol B}\Bigg\{
&
\frac12
\left(
\boldsymbol b-
\boldsymbol\Pi\boldsymbol\beta_E^*
-\frac1{a_n}\boldsymbol\kappa
\right)^{\!\top}
\boldsymbol\Theta^{-1}
\left(
\boldsymbol b-
\boldsymbol\Pi\boldsymbol\beta_E^*
-\frac1{a_n}\boldsymbol\kappa
\right)
\\
&
+
\frac12
\left(
\boldsymbol B-
\boldsymbol\Psi\boldsymbol b
-\frac1{a_n}\boldsymbol t
\right)^{\!\top}
\boldsymbol\Xi^{-1}
\left(
\boldsymbol B-
\boldsymbol\Psi\boldsymbol b
-\frac1{a_n}\boldsymbol t
\right)
+
\frac1{a_n^2}
\operatorname{Barr}_{\mathcal O}(a_n\boldsymbol B)
\Bigg\},
\end{aligned}
\]
where
\(
\operatorname{Barr}_{\mathcal O}(\boldsymbol x)
=
\sum_i
\log\left(1+\frac1{s_ix_i}\right).
\)
Then
\[
\lim_{n\to\infty}
\frac1{a_n^2}
\log
\mathbb P
\left[
\sqrt n\widehat{\mathbf B}\in\mathcal O
\,\middle|\,
\widehat{\mathbf Z}=\boldsymbol z
\right]
+
L_n
=
C_0,
\]
where \(C_0\) is a constant independent of \(\boldsymbol\beta_E^*\).

The proof is provided in Appendix A.6. Theorem 3 provides an explicit approximation to the logarithm of the normalizing constant appearing in (5). Substituting this approximation into the selective log-likelihood yields a computationally tractable objective function,

\[
-\inf_{\boldsymbol b,\boldsymbol B}\Bigg\{
\frac12
(a_n\boldsymbol b-a_n\boldsymbol\Pi\boldsymbol\beta_E^*-\boldsymbol\kappa)^\top
\boldsymbol\Theta^{-1}
(a_n\boldsymbol b-a_n\boldsymbol\Pi\boldsymbol\beta_E^*-\boldsymbol\kappa)
\]
\[
\quad+
\frac12
(a_n\boldsymbol B-a_n\boldsymbol\Psi\boldsymbol b-\boldsymbol t)^\top
\boldsymbol\Xi^{-1}
(a_n\boldsymbol B-a_n\boldsymbol\Psi\boldsymbol b-\boldsymbol t)
+
\operatorname{Barr}_{\mathcal O}(a_n\boldsymbol B)
\Bigg\},
\]
where the additive constant has been omitted since it does not depend on the parameter of interest.
For numerical implementation, it is convenient to rewrite the optimization problem on the original \(\sqrt n\)-scale. Therefore, we introduce the reparameterization
$
a_n\boldsymbol b=\sqrt n\,\boldsymbol v,
a_n\boldsymbol B=\sqrt n\,\boldsymbol V,
$
which gives the following approximation of the selective log-likelihood

\[
\begin{aligned}
&
\log
\varphi
(
\sqrt n\widehat{\boldsymbol\beta}_E;
\boldsymbol\Pi\sqrt n\boldsymbol\beta_{E,n}
+\boldsymbol\kappa,
\boldsymbol\Theta
)
\\
&
+
\inf_{\boldsymbol v,\boldsymbol V}
\Bigg\{
\frac12
(
\sqrt n\boldsymbol v
-
\sqrt n\boldsymbol\Pi\boldsymbol\beta_{E,n}
-
\boldsymbol\kappa
)^\top
\boldsymbol\Theta^{-1}
(
\sqrt n\boldsymbol v
-
\sqrt n\boldsymbol\Pi\boldsymbol\beta_{E,n}
-
\boldsymbol\kappa
)
\\
&
\qquad+
\frac12
(
\sqrt n\boldsymbol V
-
\sqrt n\boldsymbol\Psi\boldsymbol v
-
\boldsymbol t
)^\top
\boldsymbol\Xi^{-1}
(
\sqrt n\boldsymbol V
-
\sqrt n\boldsymbol\Psi\boldsymbol v
-
\boldsymbol t
)
+
\operatorname{Barr}_{\mathcal O}
(
\sqrt n\boldsymbol V
)
\Bigg\}.
\end{aligned}
\]

Once the approximate selective distribution has been characterized, statistical inference can be carried out based on this distribution. Specifically, we construct the selective estimator together with its corresponding observed information matrix. These quantities can be obtained by solving the following optimization problem.

\textbf{Theorem 4},
Consider solving the optimization problem
\begin{equation}
\hat{\boldsymbol{V}}_{\hat{\boldsymbol{\beta}}_E}^* = \arg\min_{\boldsymbol{V} \in \mathbb{R}^{\bar{d}}} \frac{1}{2} \big(\sqrt{n}\boldsymbol{V} - \boldsymbol{\Psi} \sqrt{n} \hat{\boldsymbol{\beta}}_E - \boldsymbol{t}\big)^\top \boldsymbol{\Xi}^{-1} \big(\sqrt{n}\boldsymbol{V} - \boldsymbol{\Psi} \sqrt{n} \hat{\boldsymbol{\beta}}_E - \boldsymbol{t}\big) + \operatorname{Barr}_{\mathcal{O}}(\sqrt{n}\boldsymbol{V}). \tag{7}
\end{equation}
The selective estimator and the corresponding observed information matrix are given by
\begin{align}
&\boldsymbol{\Pi}^{-1} \hat{\boldsymbol{\beta}}_E - \frac{1}{\sqrt{n}} \boldsymbol{\Pi}^{-1} \boldsymbol{\kappa} + \boldsymbol{D}_{E,E} \boldsymbol{\Psi}^\top \boldsymbol{\Xi}^{-1} \left( \boldsymbol{\Psi} \hat{\boldsymbol{\beta}}_E + \frac{1}{\sqrt{n}} \boldsymbol{t} - \hat{\boldsymbol{V}}_{\hat{\boldsymbol{\beta}}_E}^* \right), \tag{8} \\
&\boldsymbol{D}_{E,E}^{-1} \left( \boldsymbol{\Theta}^{-1} + \boldsymbol{\Psi}^\top \boldsymbol{\Xi}^{-1} \boldsymbol{\Psi} - \boldsymbol{\Psi}^\top \boldsymbol{\Xi}^{-1} \left( \boldsymbol{\Xi}^{-1} + \nabla^2 \operatorname{Barr}_{\mathcal{O}} \left( \sqrt{n} \hat{\boldsymbol{V}}_{\hat{\boldsymbol{\beta}}_E}^* \right) \right)^{-1} \boldsymbol{\Xi}^{-1} \boldsymbol{\Psi} \right)^{-1} \boldsymbol{D}_{E,E}^{-1}. \tag{9}
\end{align}

\noindent

The proof is provided in Appendix A.7. In practice, the matrices
$\boldsymbol{\Pi}$,
$\boldsymbol{\kappa}$,
$\boldsymbol{\Theta}$,
$\boldsymbol{\Psi}$,
$\boldsymbol{t}$,
and
$\boldsymbol{\Xi}$
are computed using the observed information matrix
$\widehat{\boldsymbol{\Sigma}}$.
Substituting these quantities into (8) and (9) yields the selective estimator and the corresponding observed information matrix, respectively.
The complete inference procedure is summarized in Algorithm 2.

\begin{center}
\begin{tabular}{p{14cm}}
\toprule
\textbf{Algorithm 2 : Inference Based on the Approximate Selective Distribution} \\
\midrule

1. Compute the aggregated estimator $\widehat{\boldsymbol{\beta}}_E$.\\[4pt]

2. Solve the optimization problem in (7) to obtain
$\widehat{\boldsymbol{V}}_{\widehat{\boldsymbol{\beta}}_E}^{*}$.\\[4pt]

3. Compute the selective estimator
$\widehat{\boldsymbol{\beta}}_E^{(S)}$
and the corresponding observed information matrix
$\widehat{\boldsymbol{\Sigma}}_{E,E}^{(S)}$
according to (8) and (9).\\[4pt]

4. For each selected variable $j\in[d]$, compute the standard error
\[
\widehat{\sigma}_j^{(S)}
=
\sqrt{
\left(
\widehat{\boldsymbol{\Sigma}}_{E,E}^{(S)}
\right)^{-1}_{j,j}
}.
\]

5. Compute the two-sided $p$-value for testing
$H_0:\beta_{E,j}=0$
by
\[
2\cdot
\min
\left(
\Phi
\left(
\frac{\sqrt{n}\,\widehat{\beta}_{E,j}^{(S)}}
{\widehat{\sigma}_j^{(S)}}
\right),
\,
\overline{\Phi}
\left(
\frac{\sqrt{n}\,\widehat{\beta}_{E,j}^{(S)}}
{\widehat{\sigma}_j^{(S)}}
\right)
\right),
\]
where $\overline{\Phi}=1-\Phi$ denotes the upper-tail probability of the standard normal distribution.\\[4pt]

6. Construct the two-sided $100(1-\alpha)\%$ confidence interval for
$\beta_{E,j}$ as
\[
\widehat{\beta}_{E,j}^{(S)}
\pm
z_{1-\alpha/2}
\frac{\widehat{\sigma}_j^{(S)}}{\sqrt{n}}.
\]
\\
\bottomrule
\end{tabular}
\end{center}

\section{Simulations}

Our simulation studies consider one central machine and $K$ local machines.
For each observation, the covariate vector $\boldsymbol{X}_i$ is independently generated from
$
\mathcal{N}_p(\mathbf{0},\boldsymbol{\Sigma}),
$
where $p=100$ and the covariance matrix follows an autoregressive structure,
$
\boldsymbol{\Sigma}_{ij}=0.5^{|i-j|}.
$
The true coefficient vector is sparse with $s=5$ nonzero coefficients. Each nonzero coefficient is generated independently as
$
\pm\sqrt{2c\log p},
$
with equal probability, where $c$ controls the signal strength.

The response is generated according to the linear quantile regression model
$
Y_i=\boldsymbol{X}_i^{\top}\boldsymbol{\beta}
+
\varepsilon_i.
$
To examine the finite-sample performance of the proposed procedure under
different error distributions, we consider both Gaussian and heavy-tailed
errors. Specifically, we consider a Gaussian error distribution with
variance $\sigma^2=1$ and a Student's $t$ distribution with three degrees of freedom,
scaled to have unit variance. For each quantile level, the error
distribution is appropriately shifted so that its $\tau$-th quantile
is zero. The errors are generated independently of $\boldsymbol{X}_i$. Therefore,
$
Q_{\tau}(\varepsilon_i)=0,
$
which ensures that
$
Q_{\tau}(Y_i\mid\boldsymbol{X}_i)=
\boldsymbol{X}_i^{\top}\boldsymbol{\beta}.
$
For the kernel estimation of the error density at zero, we use the
standard biweight kernel (2005),
$
G(u)=\frac{15}{16}(1-u^2)^2 I(|u|\leq 1).
$
For machine $k$, the bandwidth $h$ is set to
$
h=c_h\sqrt{\frac{s\log p}{n_k}},
$
where $c_h=1$ throughout the simulations.

Following the simulation designs of Liu and Panigrahi (2025), we consider Scenarios \textbf{(I)}-\textbf{(III)} with $\tau \in \{0.3,0.5,0.7\}$ under both normal and $t_3$ errors. We additionally include Scenario \textbf{(IV)}, which fixes $\tau=0.5$ to enable a comparison between quantile and mean regression based distributed selective inference. All experiments are repeated 500 times.

Scenario \textbf{(I)}: We vary the number of local machines
$K\in\{2,4,6,8\}$. The central machine contains $n_0=1000$ observations, and the local
sample size is set to
$n_k=\lfloor n_{\mathrm{loc}}/K\rfloor$,
where $n_{\mathrm{loc}}=8000$.
The signal strength is fixed at $c=0.7$.

Scenario \textbf{(II)}: We vary the signal strength
$c\in\{0.3,0.5,0.7,0.9\}$. We fix $K=2$. The central machine contains
$n_0=2000$ observations, and each of the two local machines contains
$n_k=4000$ observations.

Scenario \textbf{(III)}: We vary the number of observations on the
central machine,
$n_0\in\{250,500,1000,2000\}$. We fix $K=3$, and each local machine
contains $n_k=2000$ observations. The signal strength is fixed at $c=0.7$.

Scenario \textbf{(IV)}: We compare \textbf{QR-Dist-SI} with its mean-regression counterpart, \textbf{LS-Dist-SI}. We fix $\tau=0.5$ and use the same setting as in Scenario \textbf{(I)}. Both methods are applied to the same data under normal and $t_3$ errors, for which the conditional mean and median share the same coefficient vector.

We evaluate the empirical coverage probability and average length of the 90\%
confidence intervals for the selected coefficients. In Scenarios \textbf{(I)}- \textbf{(III)}, to
assess the performance of the proposed method, we compare the distributed
selective inference procedure (\textbf{Dist-SI}) with two benchmark methods, \textbf{Splitting}
and \textbf{Naive}, following the comparison schemes commonly considered in selective
inference and distributed selective inference [21,31].
In Scenario \textbf{(IV)}, we focus on the comparison between quantile and mean regression.
The proposed procedure at $\tau=0.5$ is denoted by \textbf{QR-Dist-SI}, while
\textbf{LS-Dist-SI} denotes the corresponding least-squares-based distributed selective
inference procedure. Both methods are evaluated using the same coverage and average
confidence interval length criteria for the selected coefficients as
in the preceding scenarios.

(i) \textbf{Dist-SI} performs selective inference by conditioning on the
model selection event and efficiently reuses summary statistics from all
machines after model selection.

(ii) \textbf{Splitting} performs inference using only the data stored on the
central machine after model selection, thereby avoiding the dependence
between model selection and inference.

(iii) \textbf{Naive} uses all observations for both model selection and
inference without accounting for the double use of data and the resulting
selection bias.


\begin{figure}[htbp]
    \centering
    \includegraphics[width=0.8\textwidth]{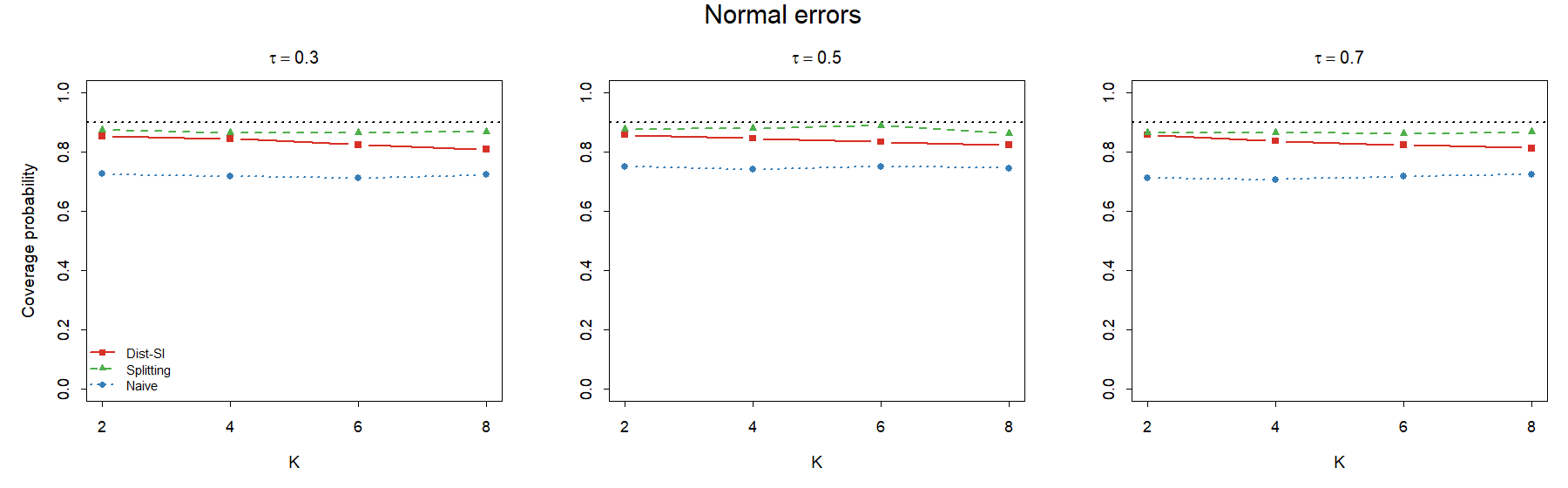}
    
    \vspace{0.3cm}
    
    \includegraphics[width=0.8\textwidth]{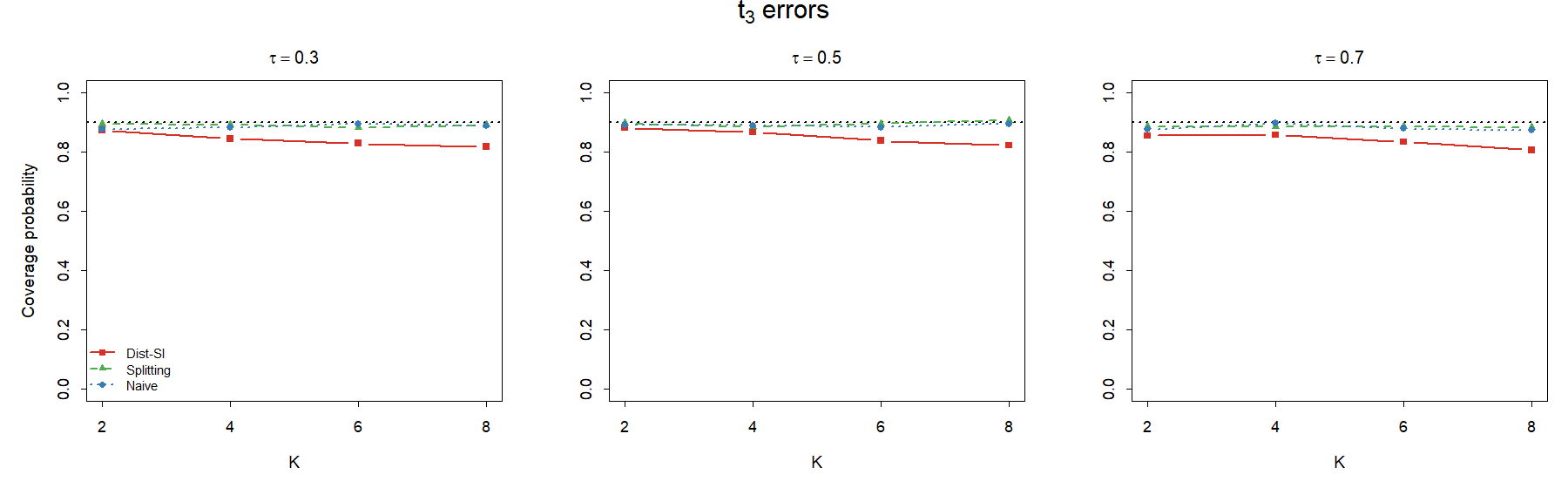}
    \caption{Coverage probabilities of the 90\% confidence intervals in Scenario 1
    under normal errors and \(t_3\) errors.}
    \label{fig:s1_coverage}
\end{figure}

\begin{figure}[htbp]
    \centering
    \includegraphics[width=0.8\textwidth]{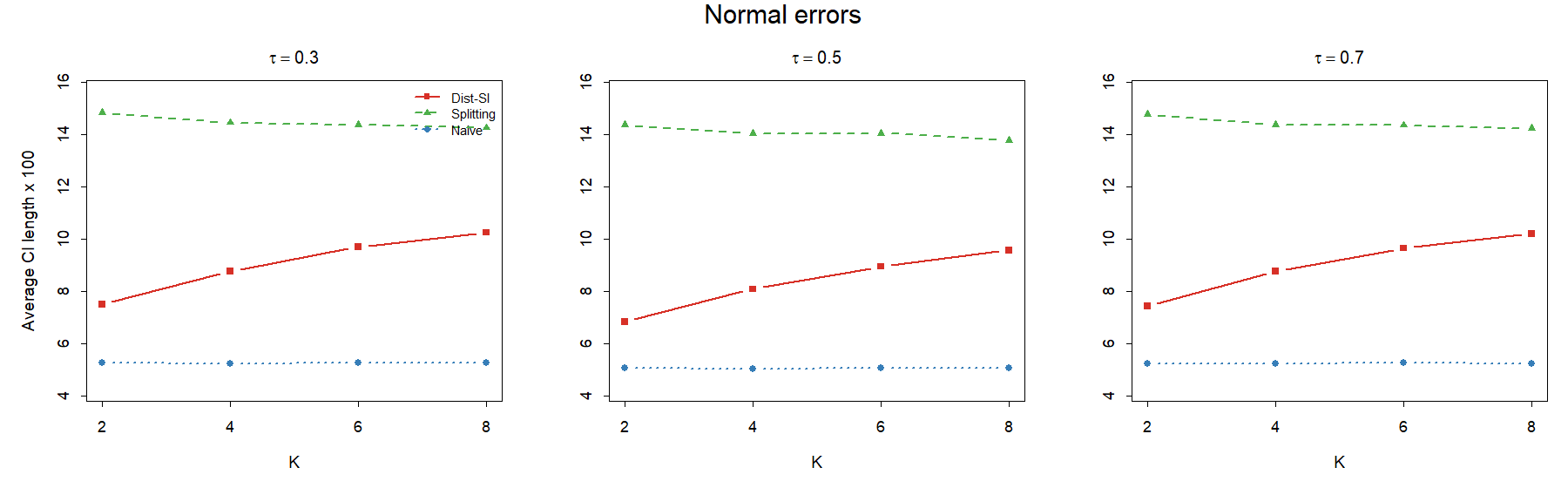}
    
    \vspace{0.3cm}
    
    \includegraphics[width=0.8\textwidth]{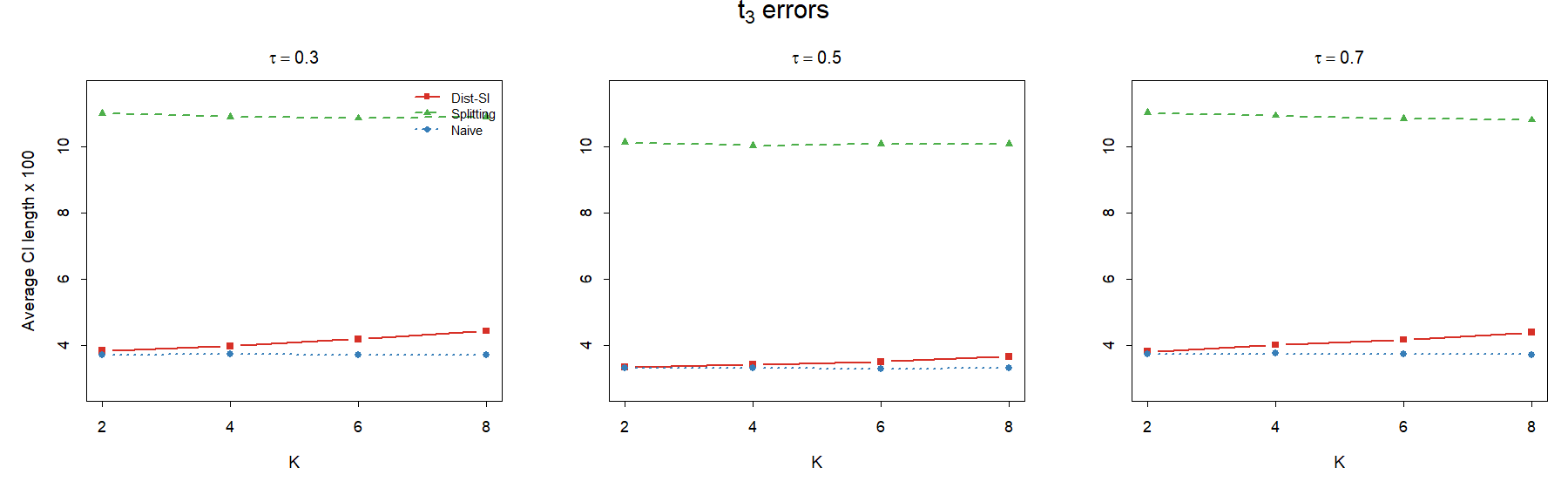}
    \caption{Average lengths of the 90\% confidence intervals in Scenario 1
    under normal errors and \(t_3\) errors.}
    \label{fig:s1_length}
\end{figure}


\begin{figure}[htbp]
    \centering
    \includegraphics[width=0.8\textwidth]{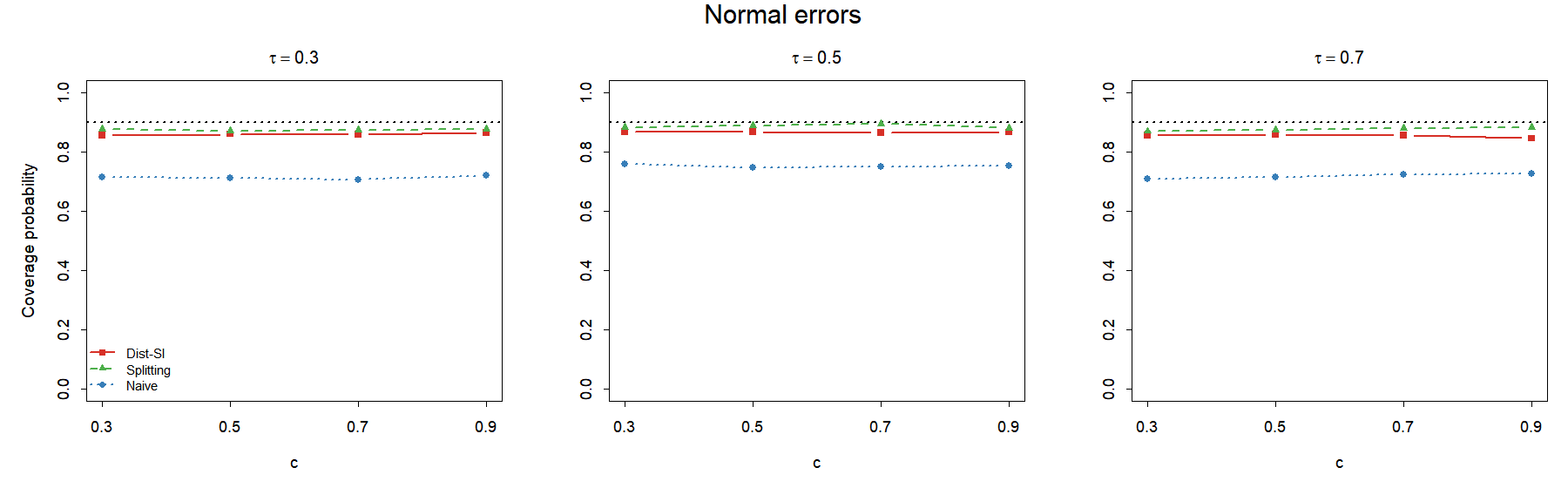}
    
    \vspace{0.3cm}
    
    \includegraphics[width=0.8\textwidth]{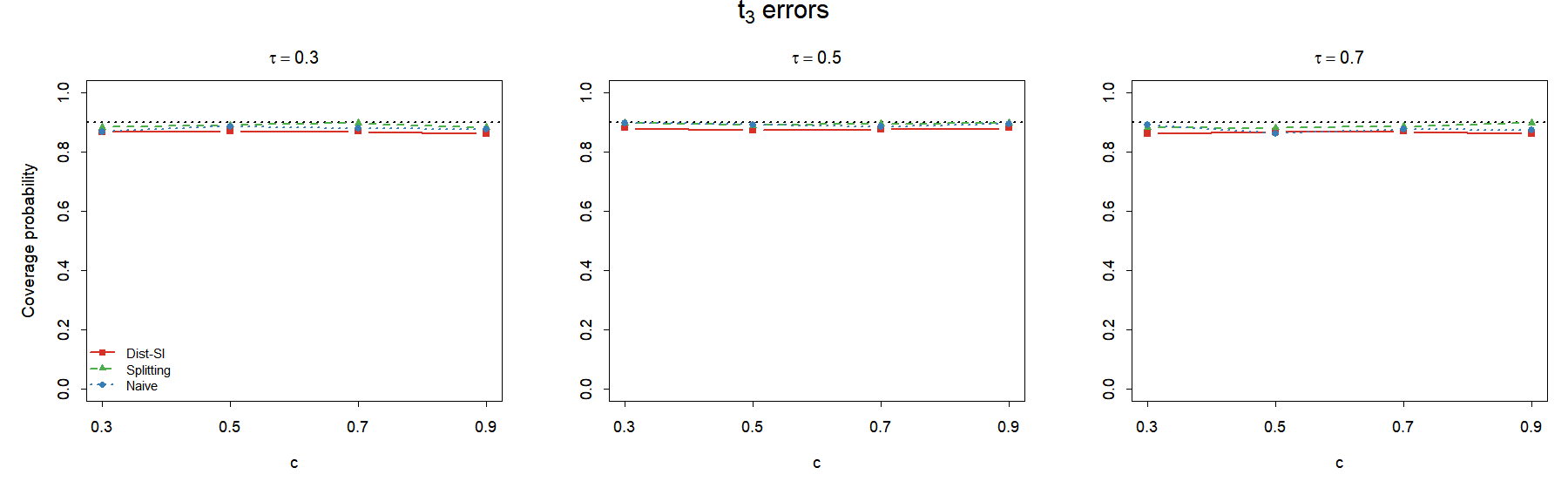}
    \caption{Coverage probabilities of the 90\% confidence intervals in Scenario 2
    under normal errors and \(t_3\) errors.}
    \label{fig:s2_coverage}
\end{figure}

\begin{figure}[htbp]
    \centering
    \includegraphics[width=0.8\textwidth]{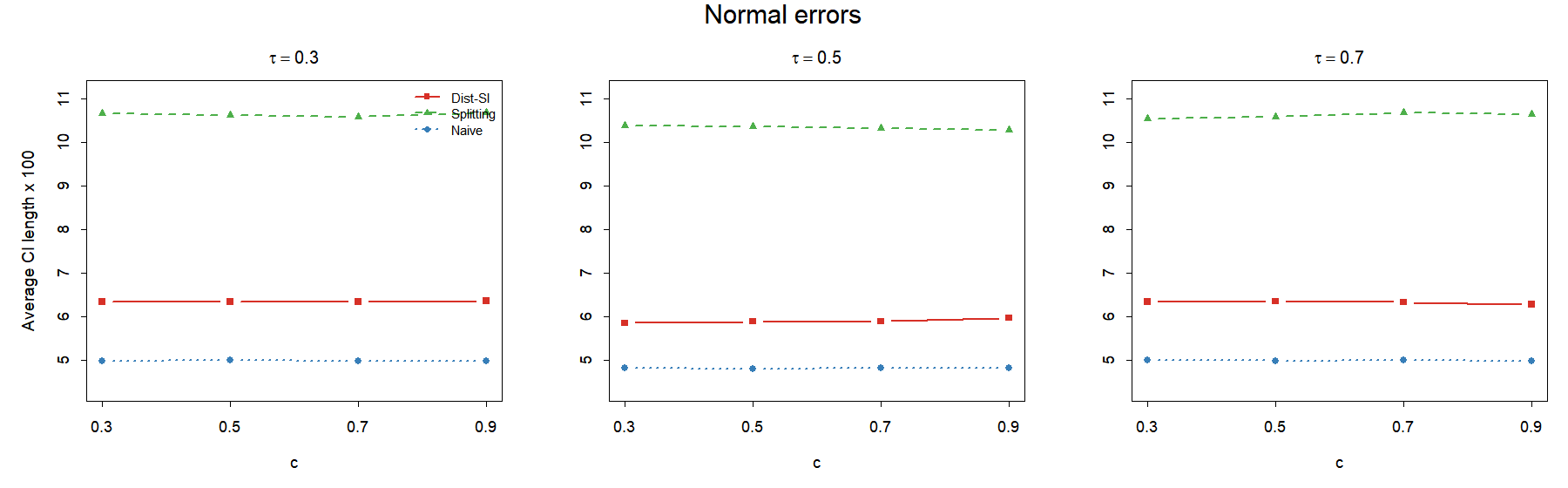}
    
    \vspace{0.3cm}
    
    \includegraphics[width=0.8\textwidth]{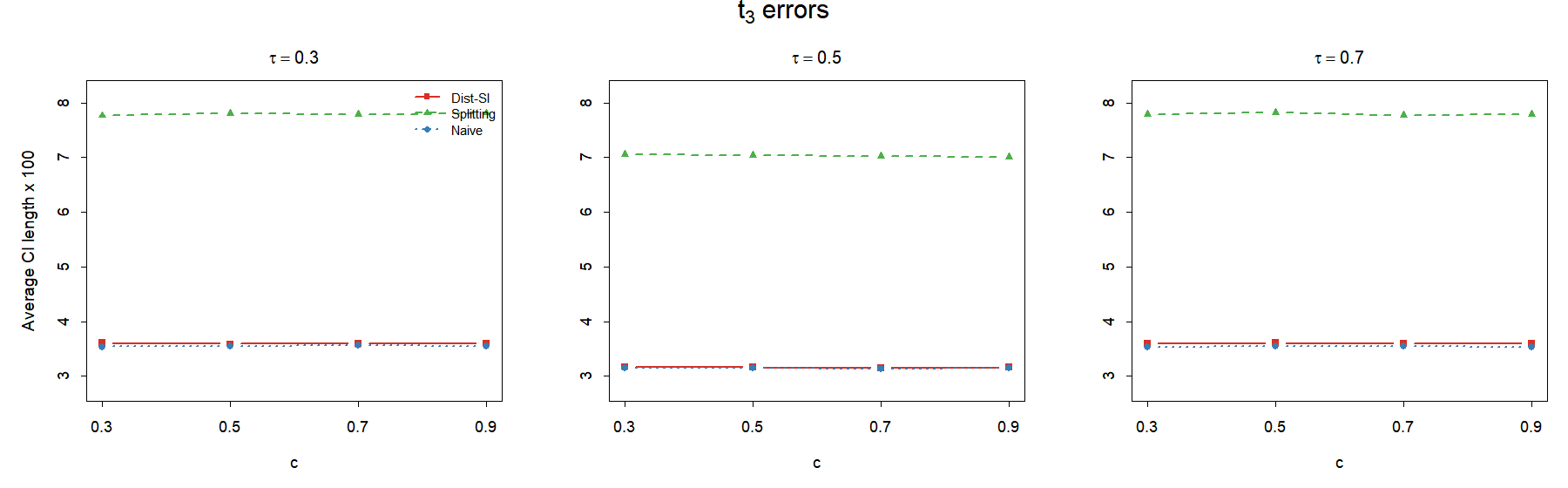}
    \caption{Average lengths of the 90\% confidence intervals in Scenario 2
    under normal errors and \(t_3\) errors.}
    \label{fig:s2_length}
\end{figure}


\begin{figure}[htbp]
    \centering
    \includegraphics[width=0.8\textwidth]{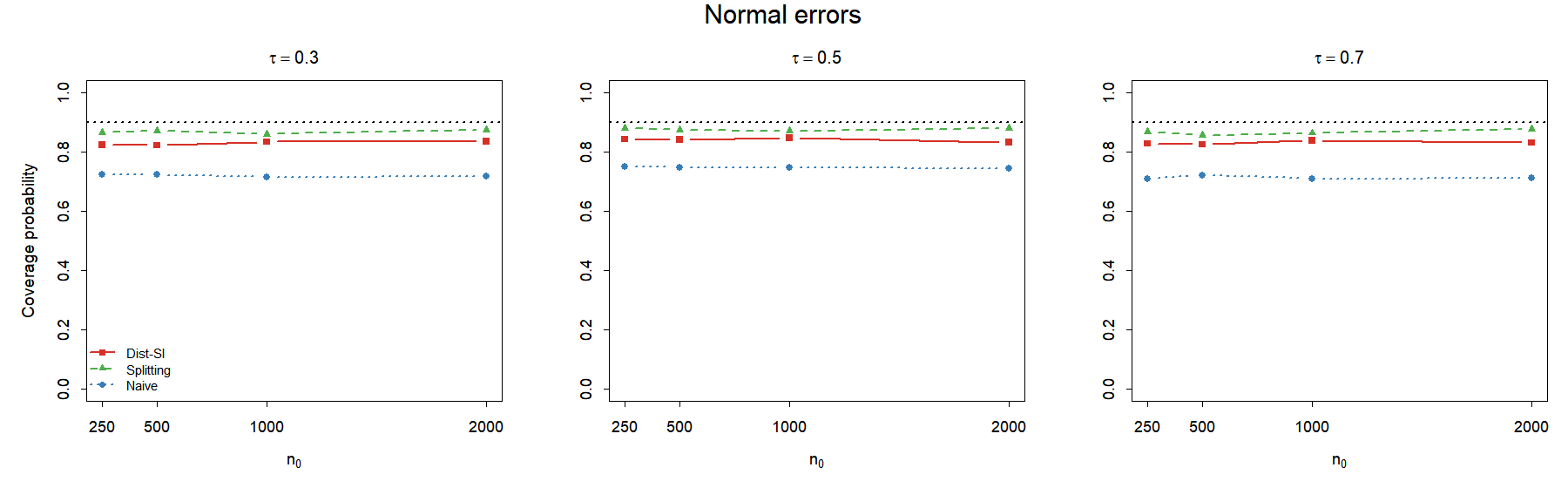}
    
    \vspace{0.3cm}
    
    \includegraphics[width=0.8\textwidth]{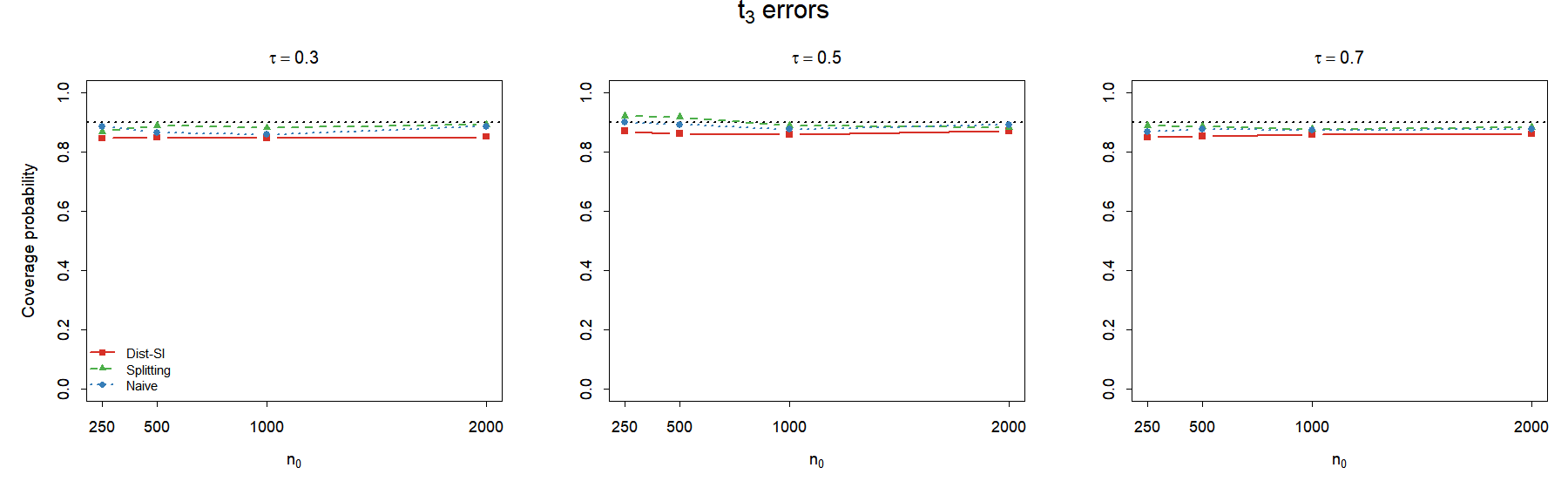}
    \caption{Coverage probabilities of the 90\% confidence intervals in Scenario 3
    under normal errors and \(t_3\) errors.}
    \label{fig:s3_coverage}
\end{figure}

\begin{figure}[htbp]
    \centering
    \includegraphics[width=0.8\textwidth]{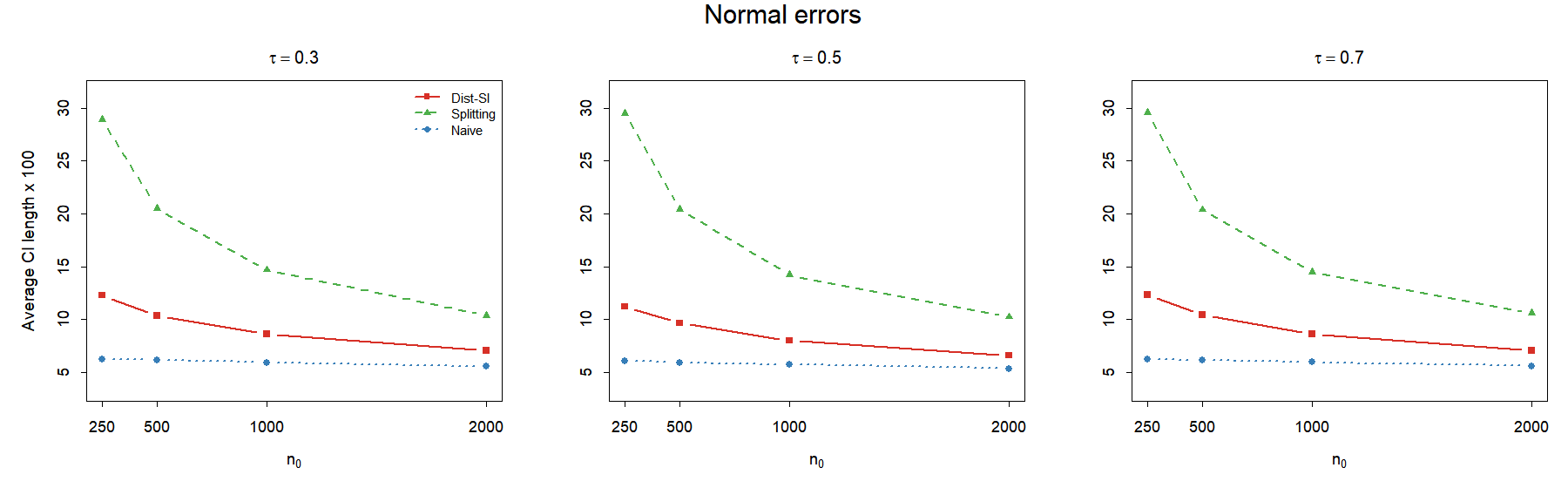}
    
    \vspace{0.3cm}
    
    \includegraphics[width=0.8\textwidth]{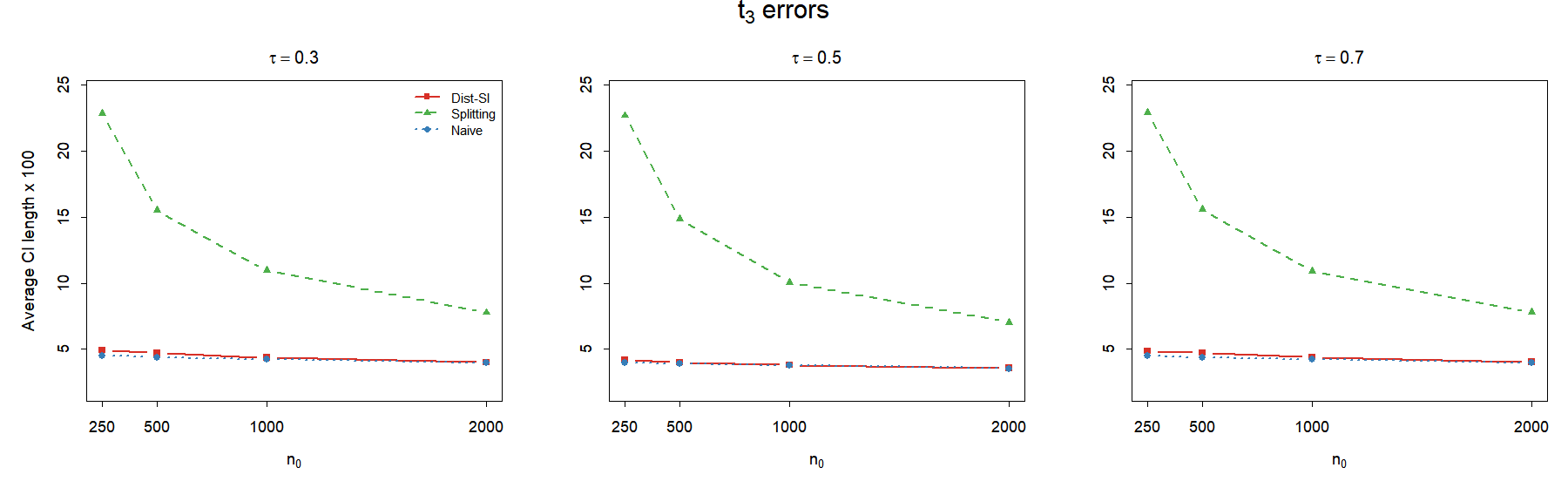}
    \caption{Average lengths of the 90\% confidence intervals in Scenario 3
    under normal errors and \(t_3\) errors.}
    \label{fig:s3_length}
\end{figure}


\begin{figure}[htbp]
    \centering
    \includegraphics[width=0.8\textwidth]{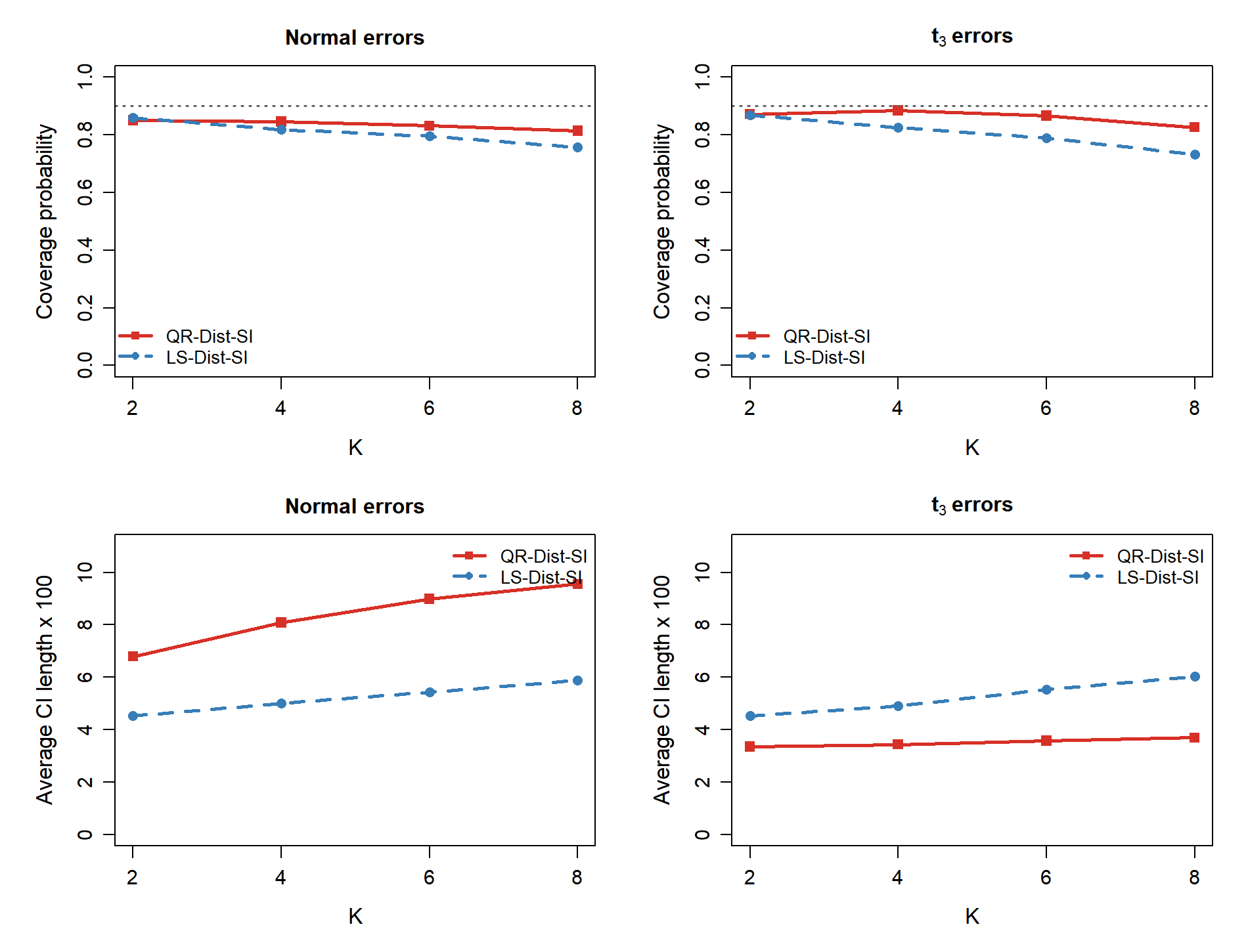}
    \caption{Coverage probabilities and average lengths of the 90\% confidence
    intervals in Scenario 4.}
    \label{fig:s4_qr_ls}
\end{figure}

The simulation results show clear differences among the inference procedures
in terms of coverage accuracy and interval length. In Scenarios \textbf{(I)}-\textbf{(III)},
\textbf{Dist-SI} generally provides higher coverage than \textbf{Naive} under normal errors,  although its coverage remains moderately below the nominal
level of 90\%. \textbf{Splitting} is typically closer to the nominal level.
Under $t_3$ errors, the coverage differences between
\textbf{Dist-SI} and \textbf{Naive} are less pronounced. The coverage of
\textbf{Dist-SI} decreases moderately as the number of local machines
increases, while remaining relatively stable across different signal
strengths, central-machine sample sizes, and quantile levels.

In terms of efficiency, \textbf{Dist-SI} consistently produces substantially
shorter confidence intervals than \textbf{Splitting} under both error
distributions. This advantage is particularly evident when the central
machine contains relatively few observations, and becomes smaller as the
central-machine sample size increases. Although \textbf{Naive} often produces
intervals comparable to or shorter than those of \textbf{Dist-SI}, it does
not account for model selection and exhibits considerable undercoverage under
normal errors. These results indicate that \textbf{Dist-SI} provides a
favorable balance between coverage accuracy and interval efficiency by
incorporating information from the local machines. The results for Scenarios \textbf{(I)--(III)} are also in line with previous work on selective inference for quantile regression and distributed selective inference .

Scenario \textbf{(IV)} examines the relative performance of the quantile-
and mean-regression-based procedures. Under normal errors,
\textbf{LS-Dist-SI} yields shorter confidence intervals than
\textbf{QR-Dist-SI}, as expected when least-squares inference is well suited
to the underlying error distribution. Under $t_3$ errors, however,
\textbf{QR-Dist-SI} produces substantially shorter intervals and generally
achieves higher coverage than \textbf{LS-Dist-SI}, with the difference
becoming more pronounced as the number of local machines increases.
These results suggest that the relative performance of the two procedures
depends on the error distribution. In particular, \textbf{LS-Dist-SI} is more
efficient under Gaussian errors, whereas \textbf{QR-Dist-SI} performs more favorably
under heavy-tailed errors.

\section{Real data analysis}

We apply the proposed distributed selective inference procedure to a
real-world aviation dataset obtained from the A320 aircraft records studied
in Jin et al.(2024). The dataset contains large-scale flight
measurements collected during aircraft operations. We take the maximum
vertical acceleration during landing as the response variable and use the
remaining flight measurements as candidate explanatory variables. The
objective of this analysis is to identify flight-related factors associated
with landing vertical acceleration and to conduct valid post-selection
inference for the selected factors in a distributed data environment.

We randomly sample approximately 10,000 observations and construct a
distributed setting consisting of one central machine and two local
machines. The central machine contains \(n_0=1000\) observations, while
each local machine contains approximately 4000 observations. We consider
the quantile levels \(\tau=0.1, 0.5,\) and \(0.9\) to investigate factors
associated with different parts of the conditional distribution of landing
vertical acceleration. We compare the proposed \textbf{Dist-SI} procedure
with the \textbf{Splitting} and \textbf{Naive} approaches, and construct
\(90\%\) confidence intervals for the selected coefficients.

Figure~\ref{fig:08} presents the confidence intervals for coefficients
rejected by at least one method. At \(\tau=0.1, 0.5,\) and \(0.9\),
\textbf{Dist-SI} identifies 11, 3, and 10 significant variables,
respectively. The selected variables vary across the quantile levels,
reflecting different factors associated with landing vertical acceleration.
Compared with \textbf{Splitting}, \textbf{Dist-SI} generally yields shorter
confidence intervals by incorporating information from the distributed
datasets while accounting for the model-selection effect. Although
\textbf{Naive} often produces shorter intervals, it ignores the uncertainty
introduced by variable selection. These results illustrate the efficiency
of \textbf{Dist-SI} for post-selection inference in distributed quantile
regression.

Overall, the real-data analysis demonstrates that \textbf{Dist-SI}
provides an effective approach for identifying flight-related factors
associated with aircraft landing vertical acceleration in a distributed
data environment. By incorporating information from the distributed
datasets while accounting for the preceding variable-selection step,
\textbf{Dist-SI} provides more efficient selection-adjusted inference than
the data-splitting approach across different conditional quantiles.
\par
\vspace{0.2cm}

\noindent
\begin{minipage}{\textwidth}
    \centering

    \includegraphics[
        width=0.70\textwidth,
        height=0.5\textheight,
        keepaspectratio
    ]{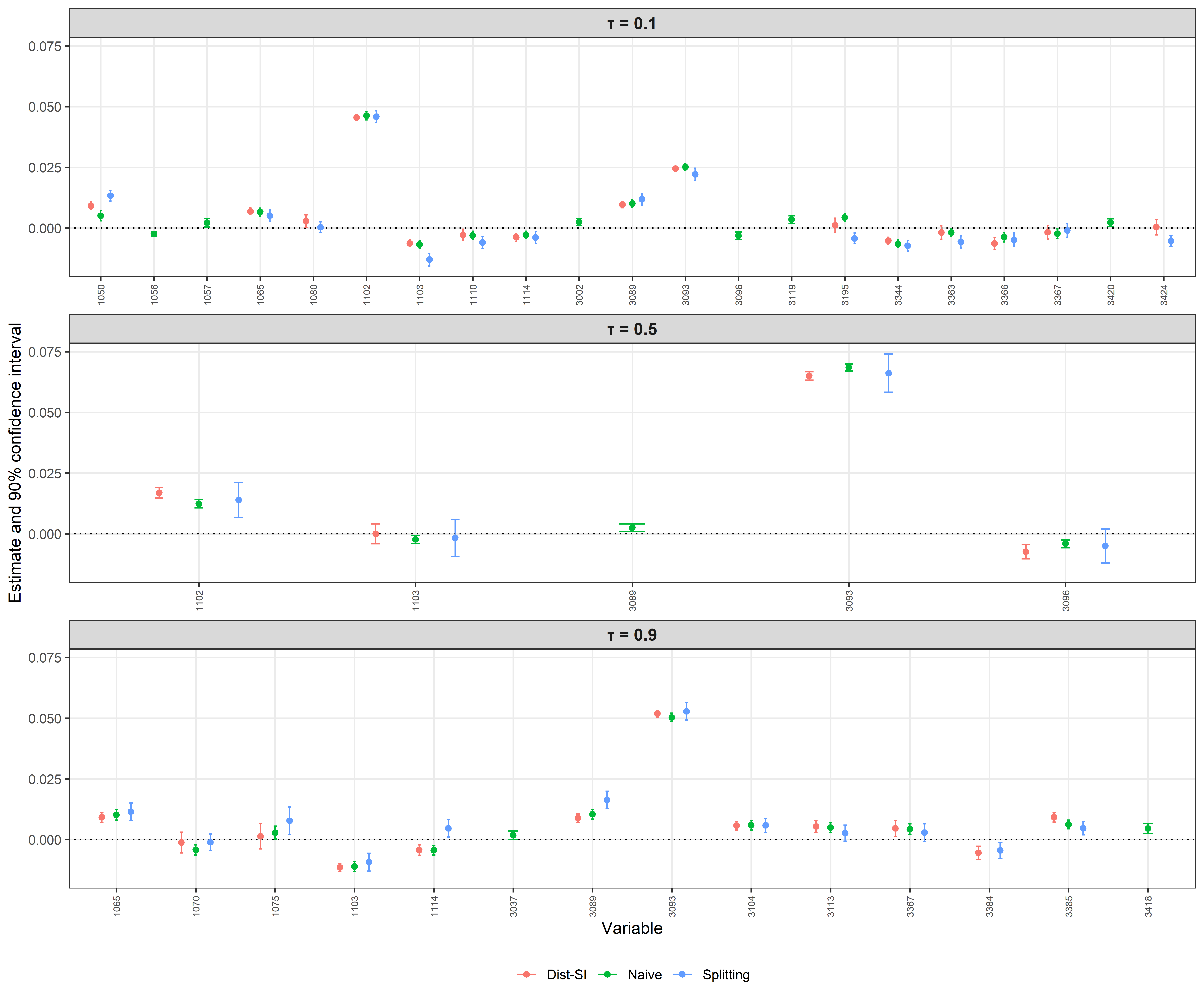}

    \captionof{figure}{Confidence intervals for coefficients rejected by at least one method at $\tau=0.1$, $0.5$, and $0.9$.}
    \label{fig:08}

\end{minipage}

\section{Discussion}

This paper develops a distributed selective inference framework for high-dimensional quantile regression. By combining surrogate response transformation with randomized selective inference, the proposed method enables valid post-selection inference while maintaining communication efficiency in distributed environments. The proposed procedure requires only three rounds of communication and establishes asymptotically valid selective confidence intervals and hypothesis tests. A tractable large-deviation approximation further avoids computationally intensive sampling from the exact selective distribution, making the method suitable for large-scale distributed applications. Numerical studies demonstrate that the proposed procedure achieves reliable finite-sample performance across a variety of settings.

Although the proposed framework focuses on $l_1$-penalized quantile regression, the general strategy provides several promising directions for future research. One natural extension is to extend the framework to composite quantile regression, which combines information across multiple quantile levels and often achieves improved statistical efficiency. More broadly, extending communication-efficient selective inference to other robust regression models and more general distributed learning problems represents an important direction for future research.\\
~\\
\textbf{Acknowledgments}\\
\\
Yuan's research is supported by the Jilin Provincial Science and Technology Development Program under Grant (No. 20250102029JC).


\newpage
\begin{thebibliography}{}

\bibitem[\protect\citeauthoryear{Battey et al.}{2018}]{Battey2018}
[1] H. Battey, J. Fan, H. Liu, J. Lu, Z. Zhu. Distributed testing and estimation under sparse high dimensional models[J].
The Annals of Statistics, 46(2018), 1352--1382.
doi:10.1214/17-AOS1587.

\bibitem[\protect\citeauthoryear{Belloni and Chernozhukov}{2011}]{Belloni2011}
[2] A. Belloni, V. Chernozhukov. $\ell_1$-Penalized quantile regression in high-dimensional sparse models[J].
The Annals of Statistics, 39(2011), 82--130.
doi:10.1214/10-AOS827.

\bibitem[\protect\citeauthoryear{Belloni et al.}{2019}]{Belloni2019}
[3] A. Belloni, V. Chernozhukov, K. Kato. Valid post-selection inference in high-dimensional approximately sparse quantile regression models[J].
Journal of the American Statistical Association, 114(2019), 749--758.
doi:10.1080/01621459.2018.1448822.

\bibitem[\protect\citeauthoryear{Chao et al.}{2025}]{Chao2025}
[4] Y. Chao, L. Huang, X. Ma. Distributed optimization for penalized regression in massive compositional data[J].
Applied Mathematical Modelling, 141(2025), 115950.
doi:10.1016/j.apm.2025.115950.

\bibitem[\protect\citeauthoryear{Chen et al.}{2026}]{Chen2026}
[5] D. Chen, R. Chen, J. Tang, H. Li. Variable selection and parameter estimation in distributed high-dimensional quantile regression with responses missing at random[J].
Journal of Systems Science and Complexity, 39(2026), 385--409.
doi:10.1007/s11424-025-4064-x.

\bibitem[\protect\citeauthoryear{Chen et al.}{2020}]{Chen2020}
[6] X. Chen, W. Liu, X. Mao, Z. Yang. Distributed high-dimensional regression under a quantile loss function[J].
Journal of Machine Learning Research, 21(2020), 1--43.

\bibitem[\protect\citeauthoryear{Danilevicz et al.}{2026}]{Danilevicz2026}
[7] I. M. Danilevicz, P. Bondon, V. A. Reisen. Adaptive LASSO quantile regression with fixed effects[J].
Applied Mathematical Modelling, 153(2026), 116600.
doi:10.1016/j.apm.2025.116600.

\bibitem[\protect\citeauthoryear{Fan and Li}{2001}]{Fan2001}
[8] J. Fan, R. Li. Variable selection via nonconcave penalized likelihood and its oracle properties[J].
Journal of the American Statistical Association, 96(2001), 1348--1360.
doi:10.1198/016214501753382273.

\bibitem[\protect\citeauthoryear{Huber}{1981}]{Huber1981}
[9] P. J. Huber. Robust Statistics[M].
John Wiley \& Sons, New York, 1981.
doi:10.1002/0471725250.

\bibitem[\protect\citeauthoryear{Jiang and Yu}{2022}]{Jiang2022}
[10] R. Jiang, K. Yu. Renewable quantile regression for streaming data sets[J].
Neurocomputing, 508(2022), 208--224.
doi:10.1016/j.neucom.2022.08.009.

\bibitem[\protect\citeauthoryear{Jin et al.}{2024}]{Jin2024}
[11] J. Jin, J. Yan, R. H. Aseltine, K. Chen. Transfer learning with large-scale quantile regression[J].
Technometrics, 66(2024), 381--393.
doi:10.1080/00401706.2023.2283982.

\bibitem[\protect\citeauthoryear{Jordan et al.}{2019}]{Jordan2019}
[12] M. I. Jordan, J. D. Lee, Y. Yang. Communication-efficient distributed statistical inference[J].
Journal of the American Statistical Association, 114(2019), 668--681.
doi:10.1080/01621459.2018.1511762.

\bibitem[\protect\citeauthoryear{Kabaila and Leeb}{2006}]{Kabaila2006}
[13] P. Kabaila, H. Leeb. On the large-sample minimal coverage probability of confidence intervals after model selection[J].
Journal of the American Statistical Association, 101(2006), 619--629.
doi:10.1198/016214505000001140.

\bibitem[\protect\citeauthoryear{Kairouz et al.}{2021}]{Kairouz2021}
[14] P. Kairouz, H. B. McMahan, B. Avent, A. Bellet, M. Bennis, A. N. Bhagoji, et al. Advances and open problems in federated learning[J].
Foundations and Trends in Machine Learning, 14(2021), 1--210.
doi:10.1561/2200000083.

\bibitem[\protect\citeauthoryear{Ke et al.}{2019}]{Ke2019}
[15] Y. Ke, S. Minsker, Z. Ren, Q. Sun, W.-X. Zhou. User-friendly covariance estimation for heavy-tailed distributions[J].
Statistical Science, 34(2019), 454--471.
doi:10.1214/19-STS710.

\bibitem[\protect\citeauthoryear{Koenker}{2005}]{Koenker2005}
[16] R. Koenker. Quantile Regression[M].
Cambridge University Press, 2005.
doi:10.1017/CBO9780511754098.

\bibitem[\protect\citeauthoryear{Koenker and Bassett}{1978}]{Koenker1978}
[17] R. Koenker, G. Bassett. Regression quantiles[J].
Econometrica, 46(1978), 33--50.
doi:10.2307/1913643.

\bibitem[\protect\citeauthoryear{Lee et al.}{2016}]{Lee2016}
[18] J. D. Lee, D. L. Sun, Y. Sun, J. E. Taylor. Exact post-selection inference, with application to the lasso[J].
The Annals of Statistics, 44(2016), 907--927.
doi:10.1214/15-AOS1371.

\bibitem[\protect\citeauthoryear{Leeb et al.}{2015}]{Leeb2015}
[19] H. Leeb, B. M. P{\"o}tscher, K. Ewald. On various confidence intervals post-model-selection[J].
Statistical Science, 30(2015), 216--227.
doi:10.1214/14-STS507.

\bibitem[\protect\citeauthoryear{Li et al.}{2020}]{Li2020}
[20] T. Li, A. K. Sahu, A. Talwalkar, V. Smith. Federated learning: Challenges, methods, and future directions[J].
IEEE Signal Processing Magazine, 37(2020), 50--60.
doi:10.1109/MSP.2020.2975749.

\bibitem[\protect\citeauthoryear{Liu and Panigrahi}{2025}]{Liu2025}
[21] S. Liu, S. Panigrahi. Selective inference with distributed data[J].
Journal of Machine Learning Research, 26(2025), 1--44.

\bibitem[\protect\citeauthoryear{Panigrahi et al.}{2024}]{Panigrahi2024}
[22] S. Panigrahi, K. Fry, J. Taylor. Exact selective inference with randomization[J].
Biometrika, 111(2024), 1109--1127.
doi:10.1093/biomet/asae019.

\bibitem[\protect\citeauthoryear{Panigrahi and Taylor}{2023}]{Panigrahi2023}
[23] S. Panigrahi, J. Taylor. Approximate selective inference via maximum likelihood[J].
Journal of the American Statistical Association, 118(2023), 2810--2820.
doi:10.1080/01621459.2022.2081568.

\bibitem[\protect\citeauthoryear{Small et al.}{2005}]{Small2005}
[24] K. A. Small, C. Winston, J. Yan. Uncovering the distribution of motorists' preferences for travel time and reliability[J].
Econometrica, 73(2005), 1367--1382.
doi:10.1111/j.1468-0262.2005.00619.x.

\bibitem[\protect\citeauthoryear{Su and Wang}{2021}]{Su2021}
[25] M. Su, W. Wang. Elastic net penalized quantile regression model[J].
Journal of Computational and Applied Mathematics, 392(2021), 113462.
doi:10.1016/j.cam.2021.113462.

\bibitem[\protect\citeauthoryear{Tian et al.}{2016}]{Tian2016}
[26] X. Tian, S. Panigrahi, J. Markovic, N. Bi, J. Taylor. Selective sampling after solving a convex problem[EB/OL].
arXiv:1609.05609, 2016.

\bibitem[\protect\citeauthoryear{Tian and Taylor}{2018}]{Tian2018}
[27] X. Tian, J. Taylor. Selective inference with a randomized response[J].
The Annals of Statistics, 46(2018), 679--710.
doi:10.1214/17-AOS1554.

\bibitem[\protect\citeauthoryear{Tibshirani}{1996}]{Tibshirani1996}
[28] R. Tibshirani. Regression shrinkage and selection via the lasso[J].
Journal of the Royal Statistical Society: Series B (Methodological), 58(1996), 267--288.
doi:10.1111/j.2517-6161.1996.tb02080.x.

\bibitem[\protect\citeauthoryear{Tibshirani et al.}{2018}]{Tibshirani2016}
[29] R. J. Tibshirani, J. Taylor, R. Lockhart, R. Tibshirani. Exact post-selection inference for sequential regression procedures[J].
Journal of the American Statistical Association, 113(2018), 576--587.
doi:10.1080/01621459.2016.1228545.

\bibitem[\protect\citeauthoryear{Wainwright}{2019}]{Wainwright2019}
[30] M. J. Wainwright. High-Dimensional Statistics: A Non-Asymptotic Viewpoint[M].
Cambridge University Press, 2019.
doi:10.1017/9781108627771.

\bibitem[\protect\citeauthoryear{Wang et al.}{2025}]{Wang2025}
[31] Y. Wang, S. Panigrahi, X. He. Asymptotically-exact selective inference for quantile regression[J].
The Annals of Statistics, 53(2025), 2356--2379.
doi:10.1214/24-AOS2488.

\bibitem[\protect\citeauthoryear{Zhang et al.}{2015}]{Zhang2015}
[32] Y. Zhang, J. Duchi, M. Wainwright. Divide and conquer kernel ridge regression: A distributed algorithm with minimax optimal rates[J].
Journal of Machine Learning Research, 16(2015), 3299--3340.

\end{thebibliography}
\end{document}